\documentclass[12pt]{article}

\usepackage{amsmath,amssymb}
\usepackage{color}

\newcommand{\be}{\begin{equation}}
\newcommand{\ee}{\end{equation}}

\renewcommand{\theequation}{\arabic{section}.\arabic{equation}}

\def\tr{{\rm tr}\,}
\def\Tr{{\rm Tr}\,}
\def\cN{{\cal N}}

\def\bea{\begin{eqnarray}}
\def\eea{\end{eqnarray}}
\def\nn{\nonumber}
\def\ui{{\underline{i}}}

\def\theequation{\arabic{section}.\arabic{equation}}

\begin{document}
\begin{titlepage}

\vspace{1cm}

\begin{center}
{\Large\bf Low-energy effective actions in \\[3mm] three-dimensional
extended SYM theories} \vspace{1cm}

{\large\bf I.L. Buchbinder$\,{}^{*}$,
N.G. Pletnev$\,{}^{+}$,
I.B. Samsonov$\,{}^{\star}$
\footnote{On leave from Tomsk Polytechnic University, 634050 Tomsk,
Russia}
\\[8pt]
\it\small $^*$Department of Theoretical Physics, Tomsk State
Pedagogical University,\\ 634061 Tomsk, Russia,
 {\tt email:\ joseph@tspu.edu.ru}
\\[8pt]
$^+$Department of Theoretical Physics, Institute of Mathematics,
630090 Novosibirsk, Russia\\
{\tt email:\ pletnev@math.nsc.ru}\\[8pt]
$^\star$INFN, Sezione di Padova, 35131 Padova, Italy\\
{\tt email:\ samsonov@mph.phtd.tpu.ru}}
\end{center}
\vspace{0.5cm}

\begin{abstract}
We develop the background field method in the $\cN=2$, $d=3$
superspace for studying effective actions in
three-dimensional SYM models which live in the world-volume
of various 2-branes.
In particular, the low-energy effective action for the $\cN=2$ quiver
gauge theory with four chiral superfields in the bifundamental representation
is studied. This gauge theory describes the D2 brane probing the conifold
singularity. Surprisingly, the leading terms in this effective
action  reproduce the classical action of the Abelian ABJM
theory confirming the fact that the M2 brane can be considered as the
effective theory for the D2 brane at strong coupling.
Apart from this $\cN=2$ quiver gauge theory we study the low-energy
effective action in pure $\cN=2$, $\cN=4$ and $\cN=8$ SYM
theories with gauge group SU$(N)$ spontaneously broken down to an
Abelian subgroup. In particular, for the $\cN=4$ SYM we find
similar correspondence between the leading terms in its effective
action and the classical action of the Abelian Gaiotto-Witten
theory.
\end{abstract}

\end{titlepage}

\tableofcontents
\newpage

\setcounter{equation}{0}
\section{Introduction}
The models living in the world-volume of 2-branes in the ten or
eleven dimensional supergravity can be described by
three-dimensional supersymmetric gauge theories with extended
supersymmetry. The well-known example is the $\cN=8$, $d=3$ SYM
theory which describes a stack of D2 branes in flat background.
Quantizing the $\cN=8$, $d=3$ SYM theory helps to understand the
effective quantum dynamics of the D2 branes.
Another very interesting example is the model of multiple M2
branes whose world-volume field theory was constructed quite
recently in the series of papers \cite{BLG,ABJM}. These
models are usually referred to as the Bagger-Lambert-Gustavsson (BLG)
or Aharony-Bergman-Jafferis-Maldacena (ABJM) theories.
It is expected that quantizing the ABJM and BLG theories will shed
some light on quantum dynamics of multiple M2 branes.

Thanks to the standard Higgs mechanism, there is natural
separation between massless and massive degrees of freedom in the
supergauge theories living on D-branes.
As a result, one can define the low-energy effective
action which depends on the massless superfields while the massive
ones appear only as the internal lines in the quantum diagrams.
Such a low-energy effective action for the massless superfields is
usually well-defined and there are powerful methods of quantum
field theory for computing it both in components and in superspace
\cite{bookBK,GGRS}.

However, for supergauge theories living on M2 branes there
is no natural separation between massive and massless superfields
and the low-energy effective action for such models is not so well
understood. Indeed, compared with the D-branes, the Higgs mechanism
for the M2 brane works in a different way. In \cite{M2D2} it is
shown that when one of the scalars in the
BLG or ABJM action develops non-vanishing vev the M2 brane turns
into D2 brane described by the $\cN=8$ SYM, up to the terms
negligible at large vev. Hence, it is natural to study the
low-energy effective action for the supergauge models on D2 branes
rather than for the ABJM theory itself. Moreover, it is well-known that the M2
brane can be considered in general as the infrared limit of the D2 brane
which is usually achieved at strong gauge coupling
(see, e.g., \cite{Schwarz,KT} for recent discussions). Therefore investigating
strong coupling limit for the low-energy effective action in the
supergauge models describing D2 branes
should help to understand some quantum aspects of M2 branes.

In the present paper we study the effective action in various
three-dimensional SYM gauge theories with matter which live in the
world-volume of D2 branes on some backgrounds. In particular, we
compute the one-loop effective action in the $\cN=8$, $d=3$ SYM
theory which corresponds to the D2 branes in flat space-time and in a
quiver $\cN=2$ SYM interacting with four bifundamental chiral
superfields which corresponds to the D2 brane on a conifold near the
singularity \cite{KT,KW,Aganagic}. For the last model we show that
the leading terms in its one-loop effective action exactly reproduce
the classical action of Abelian ABJM theory in a dualized form (when
one of the chiral superfields is dualized into the gauge
superfield). This result indicates that the ABJM-like models can be
thought as the effective quantum field theories for D2 branes in an
appropriate background. Similar correspondence is established
between the leading terms in the $\cN=4$ SYM action and dualized
classical action of the Abelian Gaiotto-Witten model \cite{GW}.

The main tool for studying the effective action for the models
under consideration is the background field method in the $\cN=2$,
$d=3$ superspace which goes alone the same lines as the background
field method in the $\cN=1$, $d=4$ superspace given in details in
\cite{GGRS}. In Section 2 we review this method and apply it to
study the low-energy effective action in pure $\cN=2$ SYM theory.
In the next section we apply the $\cN=2$ background field
method to the $\cN=4$ and $\cN=8$ SYM theories for which we
compute the one-loop effective action for the gauge group SU$(N)$
spontaneously broken down to an Abelian subgroup.
In Section 4 we study the $\cN=2$ quiver gauge theories with
two and four chiral superfields in the bifundamental representation of the
SU$(2)\times$SU$(2)$ gauge group. For the last model it is shown
that the leading terms in its one-loop effective action precisely
reproduce the classical action of Abelian ABJM theory in a dual
form. Section 5 is devoted to some discussions of the results. In
the Appendix A we consider the actions of the Gaiotto-Witten and ABJM
models in the case when one of the chiral superfields id dualized
into a gauge superfield. Some details of one-loop computations of
effective actions are collected in the Appendix B.

\setcounter{equation}{0}
\section{The $\cN=2$ background field method}
\subsection{$\cN=2$, $d=3$ SYM setup}
We follow the $\cN=2$, $d=3$ superspace conventions used in our
previous paper \cite{BPS}. In particular, the gauge covariant
derivatives
\be
\nabla_\alpha=D_\alpha+A_\alpha\,,\quad
\bar\nabla_\alpha=\bar D_\alpha+\bar A_\alpha\,,\quad
\nabla_m=\partial_m+A_m
\ee
satisfy the following algebra
\cite{HKLR,N2algebra}
\bea
\{\nabla_\alpha,\bar\nabla_\beta \}&=&-2i(\gamma^m)_{\alpha\beta}
\nabla_m +2i\varepsilon_{\alpha\beta}G\,,
\label{alg1}
\\
{}[\nabla_\alpha,\nabla_m]&=&-(\gamma_m)_{\alpha\beta}\bar
W^\beta\,,\qquad
[\bar\nabla_\alpha,\nabla_m]=(\gamma_m)_{\alpha\beta}
W^\beta\,,\label{alg2}\\
{}[\nabla_m,\nabla_n]&=&i{\rm F}_{mn}\,.
\label{algebra}
\eea
Here $G$, $W_\alpha$, $\bar W_\alpha$ and ${\rm F}_{mn}$ are
non-Abelian superfield strengths subject to the Bianchi
identities. In particular, the superfield $G$ is Hermitian
and is covariantly linear,
\be
\nabla^\alpha\nabla_\alpha G=0\,,\qquad
\bar\nabla^\alpha\bar\nabla_\alpha G=0\,.
\label{cov-lin}
\ee
The superfields $W_\alpha$ and $\bar W_\alpha$ are covariantly
(anti)chiral
\be
\nabla_\alpha\bar W_\beta=0\,,\qquad
\bar \nabla_\alpha W_\beta=0
\label{cov-chiral}
\ee
and satisfy `standard' Bianchi identity
\be
\nabla^\alpha W_\alpha=\bar\nabla^\alpha\bar W_\alpha\,.
\ee
These superfield strengths are expressed in terms of $G$ as
\be
\bar W_\alpha=\nabla_\alpha G\,,\qquad
W_\alpha=\bar\nabla_\alpha G\,.
\label{W-U}
\ee

We prefer to introduce the gauge prepotential in the
so-called chiral representation in
which the Grassmann derivative $\bar D_\alpha$ does not receive
the connection,
\be
\nabla_\alpha=e^{-2V}D_\alpha e^{2V}\,,\quad
\bar\nabla_\alpha=\bar D_\alpha \,,\quad
V^\dag=V\,.
\label{40}
\ee
In this representation the superfield strengths are expressed in
terms of the prepotential $V$ as
\be
\label{42}
G=\frac i4\bar D^\alpha(e^{-2V} D_\alpha e^{2V})\,,\quad
\bar W_\alpha=\frac i4 \nabla_\alpha \bar D^\beta (e^{-2V} D_\beta e^{2V})
\,,\quad
W_\alpha =-\frac i8 \bar D^2(e^{-2V} D_\alpha e^{2V})\,.
\ee
They are covariant under the following gauge
transformations
\be
e^{2V}\to e^{i\bar\lambda}e^{2V}e^{-i\lambda}\,,
\ee
or, in the infinitesimal form,
\be
\delta V=-\frac i2 L_V(\bar\lambda+\lambda)+\frac i2 L_V
\coth(L_V)(\bar\lambda-\lambda)\,,
\ee
where $\lambda$ and $\bar\lambda$ are chiral and antichiral
superfields, respectively, and $L_V$ denotes the commutator, e.g.,
$L_V\lambda=[V,\lambda]$.

The classical action of the $\cN=2$ SYM in the $\cN=2$ superspace
reads \be S_{\cN=2}[V]=\frac1{g^2}\tr\int d^3x d^4\theta \, G^2
=-\frac1{2g^2}\tr\int d^3x d^2\theta \, W^\alpha W_\alpha\,, \label{S1}
\ee where $g$ is the dimensionfull coupling constant, $[g]=1/2$.

\subsection{Structure of the one-loop effective action in the $\cN=2$ SYM}

Within the background field method the gauge superfield $V$ is
splitted into the `background' $V$ and `quantum' $v$ parts
\footnote{For the background gauge superfield we use the same letter as for
the original gauge superfield $V$. We hope that this will not lead to any confusions
because when the background-quantum splitting is done, the original gauge
superfield never appears in the calculation and it is not necessary to
reserve a special letter for it.}
\be
e^{2V}\to e^{2 V}e^{2gv}\,,
\label{Vv}
\ee
so that
\be
\nabla_\alpha=
e^{-2gv}{\cal D}_\alpha e^{2gv}\,,\qquad
\bar \nabla_\alpha = \bar{\cal D}_\alpha\,,
\ee
where
\be
{\cal D}_\alpha=e^{-2 V}D_\alpha e^{2 V}\,,\qquad
\bar{\cal D}_\alpha = \bar D_\alpha
\ee
are the background gauge covariant spinor derivatives.
There is a freedom in defining the gauge transformations for the
background and quantum superfields. In particular, one can
consider the so-called `background' gauge transformations
\be
e^{2 V}\rightarrow e^{i\bar\lambda}e^{2 V}e^{-i\lambda}\,,\qquad
e^{2gv}\rightarrow e^{i\tau}e^{2gv}e^{-i\tau}
\label{bg-tr}
\ee
and the `quantum' ones,
\be
e^{2 V}\rightarrow e^{2 V}\,,\qquad
e^{2gv}\rightarrow e^{i\bar\lambda}e^{2gv}e^{-i\lambda}\,.
\label{q-tr}
\ee
Here $\lambda$ and $\bar\lambda$ are (anti)chiral gauge parameters
while $\tau$ is real.

Upon such a background-quantum splitting the superfield strengths
can be decomposed in series over the coupling constant $g$,
\bea
W_\alpha&\to& W_\alpha-\frac i8\bar D^2(
2g{\cal D}_\alpha v-2g^2[v,{\cal D}_\alpha v]+O(g^3))\,,\\
G&\to&  G+\frac i2 g\bar D^\alpha{\cal D}_\alpha v
 -\frac i2 g^2\bar D^\alpha[v,{\cal D}_\alpha v]+O(g^3)\,,
\eea
where the superfield strengths ${ W}_\alpha$ and $G$
in the right hand sides are
constructed now from the background gauge superfield by the
rules (\ref{42}). The classical $\cN=2$ SYM action (\ref{S1}) can be written
as
\bea
S_{\cN=2}&\to&S_{\cN=2}[{V}]
+\frac i{g}\tr\int d^3x d^4\theta\, v{\cal D}^\alpha {W}_\alpha
+S_2[{V},v]+O(g)\,,
\label{2.21}\\
S_2[{V},v]&=&
-\tr\int d^3x d^4\theta\,v
[-\frac18 {\cal D}^\alpha \bar{\cal D}^2{\cal D}_\alpha
+i {W}^\alpha{\cal
D}_\alpha+ \frac i2 ({\cal D}^\alpha {W}_\alpha)]v\,.
\label{2.22}
\eea
The action $S_{\cN=2}[V]$ in the rhs of (\ref{2.21}) is the same as
(\ref{S1}), but it depends on the background superfield only.
In the decomposition (\ref{2.21}) we do not specify the terms with positive powers of the
coupling constant in the classical action since they are necessary
only for higher-loop computations. In the present study we restrict
ourself to the one-loop effective action which is specified by the
quadratic action $S_2$.

Within the background field method
one can usually fix the quantum gauge symmetry (\ref{q-tr}) keeping the
invariance under the background transformations. The corresponding
gauge fixing functions
\be
f= i\bar{\cal D}^2 v\,,\qquad
\bar f=i{\cal D}^2 v
\label{ff}
\ee
are defined with the help of the background-dependent covariant
spinor derivatives. These functions are covariantly (anti)chiral
and change under the quantum gauge transformations (\ref{q-tr}) as
\be
\delta f=\frac 1{2g}\bar {\cal D}^2 L_{gv}[\bar\lambda+\lambda
+\coth(L_{gv})(\lambda-\bar\lambda)]\,.
\ee
Therefore the ghost superfield action has the standard form,
\be
S_{\rm gh}=\tr\int d^3x d^4\theta\,(b+\bar b)
L_{gv}[c+\bar c+\coth (L_{gv})(c-\bar c)]
=\tr\int d^3x d^4\theta\,(\bar b c-b\bar c)+O(g)\,.
\ee

The one-loop effective action is given by the following functional
integral
\be
e^{i\Gamma_{\cN=2}[V]}=e^{iS_{\cN=2}[V]}\int {\cal D}v{\cal D}b
{\cal D}c\, \delta[f-i\bar{\cal D}^2 v]\delta[\bar f-i{\cal D}^2 v]
e^{iS_2[V,v]+iS_{\rm gh}}\,.
\ee
We average this expression with the weight
\be
1=\int {\cal D}f{\cal D}\varphi\, e^{i\tr\int d^3x d^4\theta[\frac1{8\alpha}\bar f
f-\bar\varphi\varphi]},
\ee
where $\alpha$ in the gauge-fixing parameter and
$\varphi$ is the Nielsen-Kallosh ghost. It is anticommuting and
covariantly chiral superfield. This leads to the gauge-fixing
and Nielsen-Kallosh ghost actions,
\be
S_{\rm gf}=-\frac1{16\alpha}\tr\int d^3x d^4\theta\, v\{{\cal D}^2,\bar{\cal D}^2
\}v\,,\qquad
S_{\rm NK}=-\tr\int d^3x d^4\theta \,\bar\varphi\varphi\,.
\label{S-NK}
\ee
In the Fermi-Feynman gauge $\alpha=1$ and the quadratic part of
the action with respect to the quantum superfields takes simple
form,
\be
S_2+S_{\rm gf}=-\tr\int d^3x d^4\theta
\, v \square_{\rm v} v\,,
\label{2+gf}
\ee
where
\bea
\square_{\rm v}&=&-\frac18 {\cal D}^\alpha \bar{\cal D}^2{\cal D}_\alpha
+\frac1{16}\{{\cal D}^2,\bar{\cal D}^2  \}
+\frac i2({\cal D}^\alpha{W}_\alpha)+i{W}^\alpha{\cal
D}_\alpha\nn\\
&=&{\cal D}^m {\cal D}_m +{G}^2 +i{W}^\alpha{\cal D}_\alpha
-i\bar{W}^\alpha\bar{\cal D}_\alpha
\label{v-square}
\eea
is the covariant d'Alembertian operator in the space of real
superfields. As a result, we get the following representation for the one-loop
effective action
\be
e^{i\Gamma_{\cN=2}[{V}]}=e^{iS_{\cN=2}[{V}]}\int {\cal D}v{\cal D}b
{\cal D}c{\cal D}\varphi
e^{-i\tr\int d^3x d^4\theta\, v\square_{\rm v}v+iS_{\rm gh}+iS_{\rm NK}}\,.
\label{75}
\ee
Schematically, it can be written as
\be
\Gamma_{\cN=2}=\Gamma_{\rm v}+\Gamma_{\rm ghosts}\,,\qquad
\Gamma_{\rm v}=\frac i2{\rm Tr_v}\ln \square_{\rm v}\,,\quad
\Gamma_{\rm ghosts}=-\frac{3i}2{\rm Tr}_+\ln \square_+\,.
\label{2.31}
\ee
The contribution $\Gamma_{\rm v}$ to the one-loop effective action
comes from the quantum gauge superfield while $\Gamma_{\rm
ghosts}$ is due to ghosts.
Here ${\rm Tr_v}$
and ${\rm Tr}_+$ are the functional traces of the operators acting in
the spaces of real and chiral superfields, respectively.
The operator $\square_+$ is the covariant d'Alembertian operator
acting in the space of covariantly chiral superfields
which was introduced in \cite{BPS},
\be
\square_+={\cal D}^m{\cal D}_m+G^2+\frac i2({\cal D}^\alpha {W}_\alpha)
+i{W}^\alpha {\cal D}_\alpha\,.
\label{square+}
\ee
The
explicit expressions for the traces of these operators can be
found after one specifies the gauge group and the corresponding
background gauge superfield. Further we consider one example when
the gauge group is SU$(N)$ although the other simple Lie groups
can be studied in a similar way.

\subsection{Effective action for the SU$(N)$ gauge group}
\label{sect2.3}
We will be interested in the low-energy effective action
which is a functional for the massless fields obtained by
integrating out all massive fields in a functional integral. In
gauge theories the separation between massless and massive
fields appears usually through the Higgs mechanism.
In general, the gauge group SU$(N)$ is spontaneously broken down to its
maximal Abelian subgroup, U$(1)^{N-1}$. However, in particular
cases a bigger subgroup of SU$(N)$ can be unbroken. Physically
interesting to consider minimal gauge symmetry breaking,
${\rm SU}(N)\to {\rm SU}(N-1)\times {\rm U}(1)$ because,
from the point of view of D-branes, the corresponding
effective action contains the potential which appears when
one separates one D-brane from the stack. In this section we will
consider first the general case when the gauge group is broken down to
the maximal torus and then comment on the effective action with minimal
gauge symmetry breaking.

The Lie algebra su$(N)$ consists of Hermitian traceless
matrices. Any element $v$ of su$(N)$ can be represented by a
decomposition over the Cartan-Weil basis in the gl$(N)$ algebra,
\be
(e_{IJ})_{LK}=\delta_{IL}\delta_{JK}\,,\qquad
v=\sum_{I<J}^N (v_{IJ} e_{IJ}+\bar v_{IJ}e_{JI})
+\sum_{I=1}^N v_I e_{II}\,,\quad \bar v_I=v_I\,,\quad \sum_{I=1}^N v_I=0\,.
\label{basis}
\ee
The background gauge superfield $V$ belongs to the Cartan
subalgebra spanned on $e_{II}$,
\be
V=\sum_{I=1}^N {\bf V}_I e_{II}={\rm diag}({\bf V}_1,{\bf V}_2,
\ldots,{\bf V}_N)\,,\quad
\bar {\bf V}_I={\bf V}_I\,,\quad \sum_{I=1}^N {\bf
V}_I=0 \,.
\label{V-back}
\ee
In what follows we will denote by boldface Latin letters the
matrix elements of the background superfields. In particular, each
matrix element ${\bf V}_I$ of $V$
has superfield strength ${\bf G}_I=\frac i2\bar D^\alpha D_\alpha {\bf
V}_I$ which is computed as in the Abelian case.
We will use also the following notations
\be
{\bf V}_{IJ}={\bf V}_I-{\bf V}_J\,,\quad
{\bf G}_{IJ}={\bf G}_I-{\bf G}_J\,,\quad
{\bf W}_{IJ\,\alpha}={\bf W}_{I\,\alpha}-{\bf W}_{J\,\alpha}\,.
\ee

Now we can do the matrix trace in the action (\ref{2+gf}),
\be
S_2+S_{\rm gf}=-2\sum_{I<J}^N\int d^3x d^4\theta\,
v_{IJ}\hat\square_{{\rm v}\,IJ}\, \bar v_{IJ}\,,
\ee
where
$\hat\square_{{\rm v}\,IJ}$ is the Abelian version of the operator
(\ref{v-square}) which is constructed from the Abelian gauge
superfield ${\bf V}_{IJ}$ and its superfield strengths,
\be
\hat\square_{{\rm v}\,IJ} ={\cal D}^m {\cal D}_m +{\bf G}_{IJ}^2 +i{\bf
W}_{IJ}^\alpha{\cal D}_\alpha -i\bar{\bf W}_{IJ}^\alpha\bar{\cal
D}_\alpha\,.
\label{v-square-ij}
\ee
Therefore the effective action
$\Gamma_{\rm v}$ can be written as
\be
\Gamma_{\rm v}=i\sum_{I<J}^N {\rm Tr_v} \ln \hat\square_{{\rm v}\,IJ}\,,
\label{Gamma-ij}
\ee
where ${\rm Tr_v}$ means now only the functional trace in the space of real
superfields.

In a similar way one can analyze the contributions from the ghost
superfields. Consider, for instance, the action for the
Nielsen-Kallosh ghost (\ref{S-NK}) in which the chiral superfields
are expanded over the basis (\ref{basis}) as \be \varphi=\sum_{I\ne
J}^N e_{IJ} \varphi_{IJ}\,,\qquad \bar\varphi=\sum_{I\ne J}^N
e_{IJ}\bar \varphi_{IJ}\,. \label{q-varphi} \ee Here we omit the
diagonal components because they do not interact with the background
gauge superfield (\ref{V-back}) and do not contribute to the
effective action. Then the matrix trace in the action (\ref{S-NK})
is done,
\be S_{\rm NK}=-\sum_{I\ne J}^N\int d^3x d^4\theta\,
\bar\varphi_{IJ}\varphi_{IJ}\,, \ee
where the superfields
$\bar\varphi_{IJ}$ are covariantly antichiral, \be e^{-2{\bf
V}_{IJ}}D_\alpha e^{2{\bf V}_{IJ}}\bar \varphi_{IJ}=0\ \mbox{ for
}I<J\,,\qquad e^{2{\bf V}_{IJ}}D_\alpha e^{-2{\bf V}_{IJ}}\bar
\varphi_{IJ}=0\ \mbox{ for }I>J\,. \ee We see that the chiral
superfields appear in pairs with positive and negative charges with
respect to the Abelian gauge superfield ${\bf V}_{IJ}$. This
prevents the generation of the Chern-Simons term in the one-loop
computations (there is no parity anomaly \cite{anomaly}, see also
\cite{BPS} for recent superspace calculations). As a result, the
effective action for the ghost superfields reads
\be \Gamma_{\rm
ghosts}=-3i\sum_{I<J}^N{\rm Tr}_+ \ln \hat\square_{+IJ}\,, \label{Gamma+ij}
\ee
where $\hat\square_{+IJ}$ is the Abelian version of the operator
(\ref{square+}) constructed from the gauge superfield ${\bf
V}_{IJ}$,
\be
\hat\square_{+IJ}={\cal D}^m{\cal D}_m+{\bf G}_{IJ}^2
+\frac i2({\cal D}^\alpha {\bf W}_{IJ\, \alpha}) +i{\bf
W}^\alpha_{IJ} {\cal D}_\alpha\,,
\label{square-ij}
\ee
and ${\rm Tr}_+$ denotes the functional trace in the space of chiral superfields.

To do the explicit quantum computations of traces of logarithms in
(\ref{Gamma-ij}) and (\ref{Gamma+ij}) we have to specify the constraints on the
background Abelian superfields:\\
(i) The matrix components of the background gauge superfield ${\bf
V}_{IJ}$ obey the $\cN=2$ supersymmetric Maxwell equations,
\be
D^\alpha{\bf W}_{IJ\,\alpha}
=\bar D^\alpha\bar {\bf W}_{IJ\,\alpha}=0\,.
\label{approx-1}
\ee
(ii) We study the effective action in the
so-called long-wave approximation in which the space-time
derivatives of the background are neglected,
\be
\partial_m{\bf G}_{IJ}=\partial_m {\bf W}_{IJ\,\alpha}
=\partial_m \bar {\bf W}_{IJ\,\alpha}=0\,.
\label{approx-2}
\ee
For such a background the heat kernels of the operators
(\ref{v-square-ij}) and (\ref{square-ij}) are known, see
\cite{BPS}. In the present notations they read
\bea
K_{{\rm v}\,IJ}(z,z'|s)
&=&\frac1{8(i\pi s)^{3/2}}\frac{s{\bf B}_{IJ}}{\sinh(s{\bf B}_{IJ})}
e^{is{\bf G}_{IJ}^2}
e^{\frac i4({\bf F}_{IJ}\coth s{\bf F}_{IJ})_{mn}\zeta^m(s)\zeta^n(s)}
\zeta^2(s)\bar\zeta^2(s)\,,\\
K_{+IJ}(z,z'|s)&=&-\frac14\bar{\cal D}^2 K_{{\rm
v}IJ}(z,z'|s)\,,
\eea
where ${\bf B}_{IJ}^2=\frac12 D_\alpha{\bf W}_{IJ}^\beta D_\beta
{\bf W}_{IJ}^\alpha$.
In fact, for the one-loop computations we need these
expressions only at coincident superspace points,
\bea
K_{{\rm v}\,IJ}(s)&\equiv&K_{{\rm v}\,IJ}(z,z|s)=\frac1{(i\pi)^{3/2}}\frac1{\sqrt s}
\frac{{\bf W}_{IJ}^2\bar{\bf W}_{IJ}^2}{{\bf B}_{IJ}^3}e^{is{\bf G}_{IJ}^2}
\tanh\frac{s{\bf B}_{IJ}}{2}\sinh^2\frac{s{\bf B}_{IJ}}{2}\,,\\
K_{+IJ}(s)&\equiv&K_{+IJ}(z,z|s)=\frac1{8(i\pi s)^{3/2}}s^2{\bf W}_{IJ}^2
e^{is{\bf G}_{IJ}^2}
\frac{\tanh(s{\bf B}_{IJ}/2)}{s{\bf B}_{IJ}/2}\,.
\eea
The corresponding contributions to the effective action from these heat
kernels are given by
\be
\Gamma_{\rm v}=-i\sum_{I<J}^N
\int_0^\infty\frac{ds}s\int d^3x d^4\theta\,K_{{\rm v}\,IJ}(s)\,,\qquad
\Gamma_{\rm ghosts}=-3i\sum_{I<J}^N
\int_0^\infty\frac{ds}s\int d^3x d^2\theta\,K_{+IJ}(s)\,,
\ee
or, explicitly,
\bea
\Gamma_{\rm v}&=&-\frac1\pi\sum_{I<J}^N
\int d^3x d^4\theta\int_0^\infty\frac{ds}{s\sqrt{i\pi s}}
\frac{{\bf W}_{IJ}^2\bar{\bf W}_{IJ}^2}{{\bf B}_{IJ}^3}e^{is{\bf G}_{IJ}^2}
\tanh\frac{s{\bf B}_{IJ}}{2}\sinh^2\frac{s{\bf B}_{IJ}}{2}\,,
\label{Gv}\\
\Gamma_{\rm ghosts}&=&-\frac{3}{2\pi}\sum_{I<J}^N
\int d^3x d^4\theta\bigg[{\bf G}_{IJ}\ln{\bf G}_{IJ}\nn\\&&
+\frac14\int_0^\infty\frac{ds}{\sqrt{i\pi s}}
e^{is{\bf G}_{IJ}^2}\frac{{\bf W}_{IJ}^2\bar{\bf W}_{IJ}^2}{{\bf B}_{IJ}^2}\left(
\frac{\tanh(s{\bf B}_{IJ}/2)}{s{\bf B}_{IJ}/2}-1
\right)\bigg],\label{Gh}
\eea
where in the expression for $\Gamma_{\rm ghosts}$ we restored the
full superspace measure. The sum of the expressions (\ref{Gv}) and (\ref{Gh})
gives us the resulting one-loop effective action in the pure
$\cN=2$ SYM theory for the gauge group SU$(N)$ spontaneously
broken down to U$(1)^{N-1}$. We point out
that only the leading ${\bf G}\ln\bf G$ term  in the $\cN=2$ SYM
effective action was obtained in \cite{deBoer}
using the duality transformations while the explicit quantum
computations allow us to find all higher-order $F^{2n}$ terms
encoded in the proper-time integrals (\ref{Gv}) and (\ref{Gh}).

In conclusion of this section let us comment on the case of minimal
gauge symmetry breaking ${\rm SU}(N)\to {\rm SU}(N-1)\times {\rm
U}(1)$. In this case it is convenient to choose the background gauge
superfield in the following form
\be
V=\frac1{N}{\rm
diag}\left((N-1){\bf V}, \underbrace{-{\bf V},\ldots,-{\bf
V}}_{N-1}\right),
\label{V-back1}
\ee
where $\bf V$ is Abelian gauge
superfield with the superfield strengths $\bf G$, ${\bf W}_\alpha$ and
$\bar{\bf W}_\alpha$. One can easily repeat all the above
considerations for such a background or just extract the answer
from (\ref{Gv}) and (\ref{Gh}) by substituting the corresponding
expressions for ${\bf V}_{IJ}$. For simplicity, we give here only
two leading terms in the corresponding effective action
\be
\Gamma_{\cN=2}= -\frac{3(N-1)}{2\pi}\int d^3x d^4\theta\,{\bf
G}\ln{\bf G} +\frac{9(N-1)}{128\pi}\int d^3x d^4\theta\frac{{\bf
W}^2\bar{\bf W}^2}{{\bf G}^5} +\ldots\,.
\label{G-N2}
\ee
The first
term in the rhs is responsible for the $\cN=2$ supersymmetric (and
superconformal) generalization of the Maxwell $F^2$ term while the
second one gives $F^4$ among other components. The dots here stand
for the higher orders of the Maxwell field strength.

\setcounter{equation}{0}
\section{The one-loop effective actions in the $\cN=4$ and $\cN=8$ SYM}
\subsection{Low-energy effective action in $\cN=4$ SYM}
The classical action of $\cN=4$ SYM is given by
\be
S_{\cN=4}=\frac1{g^2}\tr\int d^3x d^4\theta \, [G^2
-\frac12 e^{-2V}\bar\Phi e^{2V} \Phi]\,,
\label{N4}
\ee
where $\Phi$ is the chiral superfield in the adjoint
representation of the gauge group.
This action is invariant under the following non-Abelian gauge transformations
\be
\Phi\to e^{i\lambda} \Phi e^{-i\lambda}\,,\quad
\bar\Phi \to e^{i\bar\lambda} \bar\Phi e^{-i\bar\lambda}\,,\quad
e^{2V}\to e^{i\bar\lambda} e^{2V} e^{-i\lambda}\,,
\ee
with $\lambda$ and $\bar\lambda$ being (anti)chiral superfield gauge
parameters.

It is convenient to introduce the covariantly (anti)chiral
superfields,
\be
\bar\Phi_c=e^{-2V}\bar\Phi e^{2V}\,,\quad
\Phi_c=\Phi\,,\quad
\nabla_\alpha \bar\Phi_c=0\,,\quad
\bar\nabla_\alpha\Phi_c=0\,.
\ee
In terms of these superfields the transformations of hidden
$\cN=2$ supersymmetry are given by
\be
e^{2V}\delta_\epsilon e^{2V}=\theta^\alpha\epsilon_\alpha\bar\Phi_c
-\bar\theta^\alpha\bar\epsilon_\alpha\Phi_c\,,\quad
\delta_\epsilon\Phi_c=-i\epsilon^\alpha \bar \nabla_\alpha G\,,\quad
\delta_\epsilon\bar\Phi_c=-i\bar\epsilon^\alpha \nabla_\alpha G\,.
\label{SUSY}
\ee
Here $\epsilon_\alpha$ is the anticommuting complex parameter.
We omit the label `$c$', $\bar\Phi_c\to\bar\Phi$,
$\Phi_c\to \Phi$, adopting
that we deal with the covariantly (anti)chiral superfields in what
follows.

The generalization of the $\cN=2$ background field method to the
$\cN=4$ case is straightforward. The background-quantum splitting
of the gauge superfield $V$ in (\ref{Vv}) is extended by the
corresponding splitting for $\Phi$ and $\bar\Phi$,
\be
\Phi\to \Phi+g\phi\,,\qquad
\bar\Phi\to \bar\Phi+g\bar\phi\,.
\label{Phi-phi}
\ee
Here the superfields $\Phi$, $\phi$ and $\bar\Phi$, $\bar\phi$
in the right hand sides are covariantly (anti)chiral
with respect to the background gauge covariant
derivatives, ${\cal D}_\alpha\bar\Phi={\cal D}_\alpha\bar\phi=0$,
$\bar{\cal D}_\alpha\Phi=\bar{\cal D}_\alpha\phi=0$.
The quantum gauge transformations for these
superfields read
\be
\delta \phi=i[\lambda,\frac1g\Phi+\phi]\,,\quad
\delta\bar\phi=i[\bar\lambda,\frac1g\bar\Phi+\bar\phi]\,,\quad
\delta\Phi=\delta\bar\Phi=0\,.
\label{q-tr1}
\ee

Upon the background-quantum splitting (\ref{Vv}) and (\ref{Phi-phi}), the
$\cN=4$ SYM action (\ref{N4}) can be expanded in the series over the quantum
superfields. In particular, for the one-loop computations we need
the quadratic part of this action,
\bea
S_2&=&
-\tr\int d^3x d^4\theta\,v
[-\frac18 {\cal D}^\alpha \bar{\cal D}^2{\cal D}_\alpha
+\frac i2({\cal D}^\alpha W_\alpha)+i W^\alpha{\cal
D}_\alpha+\Phi\bar\Phi]v\nn\\
&&-\tr\int d^3xd^4\theta(-\bar\phi[\Phi,v]+\phi[\bar\Phi,v]
+\frac12\phi\bar\phi)\,.
\label{3.7}
\eea
This action is invariant under the quantum gauge transformations
(\ref{q-tr}) and (\ref{q-tr1}). Therefore we fix the quantum gauge
symmetry by the following gauge-fixing functions
\be
f= i\bar{\cal D}^2 v-\frac i2[\Phi,\bar
{\cal D}^2\square_-^{-1}\bar\phi]\,,\qquad
\bar f=i{\cal D}^2 v+\frac i2[\bar\Phi,{\cal D}^2\square_+^{-1}\phi]
\,.
\label{f-f}
\ee
In comparison with (\ref{ff}) these functions have the terms
depending on the background (anti)chiral superfields $\Phi$ and
$\bar\Phi$ which are necessary to remove the mixed terms between
the quantum gauge $v$ and (anti)chiral $\bar\phi$, $\phi$
superfields. Such a gauge fixing is usually referred to as the
generalized $R_\xi$ gauge \cite{BBP,Ovrut}.
The corresponding gauge-fixing action reads
\bea
S_{\rm gf}&=&\frac 18\tr\int d^3x d^4\theta \,\bar f f
=\frac 1{8}\tr\int d^3x d^4\theta\bigg(-\frac12 v\{{\cal D}^2,\bar{\cal D}^2 \}v
-\frac12 v\bar{\cal D}^2[\bar\Phi,{\cal D}^2\square_+^{-1}\phi]
\nn\\&&
+\frac12 v{\cal D}^2[\Phi,\bar{\cal D}^2\square_-^{-1}\bar\phi]
+\frac14[\Phi,\bar{\cal D}^2\square_-^{-1}\bar\phi]
[\bar\Phi,{\cal D}^2\square_+^{-1}\phi]
\bigg)\,.
\label{3.9}
\eea

It is convenient at this point to specify the constraints on the
background chiral superfields $\Phi$ and $\bar\Phi$,
\be
{\cal D}_\alpha\Phi=0\,,\qquad
\bar{\cal D}_\alpha\bar\Phi=0\,,
\label{approx-3}
\ee
i.e. they are covariantly constant.
For such a background the action (\ref{3.9}) simplifies,
\be
S_{\rm gf}=\tr\int d^3x d^4\theta\left(-\frac1{16} v\{{\cal D}^2,\bar{\cal D}^2 \}v
- v [\bar\Phi,\phi]
+ v[\Phi,\bar\phi]
+\frac12[\Phi,\square_-^{-1}\bar\phi]
[\bar\Phi,\phi]
\right)\,.
\ee
As a result, the quadratic part of the action for the quantum
superfields becomes very simple,
\be
S_2+S_{\rm gf}=-\tr\int d^3x d^4\theta\left[
v(\square_{\rm v}+\bar\Phi\Phi)v+
\frac12\phi(1+\bar\Phi\Phi\square_-^{-1})\bar\phi
\right].
\ee
Here we denote $\bar\Phi\Phi v=[\bar\Phi,[\Phi, v]]$ and similar
$\bar\Phi\Phi \bar\phi=[\bar\Phi,[\Phi,\bar\phi]]$.

The quantum gauge transformations (\ref{q-tr}) and (\ref{q-tr1})
define the action for the ghost superfields,
\bea
S_{\rm gh}&=&\tr\int d^3x d^4\theta(b+\bar b)
L_{gv}[c+\bar c+\coth (L_{gv})(c-\bar c)]\nn\\&&
-\tr\int d^3x d^4\theta \left(
b[\Phi,\square_-^{-1}[\bar\Phi+g\bar\phi, \bar c]]
-\bar b[\bar \Phi,\square_+^{-1}[\Phi+g\phi,c]]
\right)\nn\\
&&-\tr\int d^3x d^4\theta\,\bar\varphi\varphi\,. \eea Here $b$ and
$c$ are standard Faddeev-Popov ghosts while $\varphi$ is the
Nielsen-Kallosh ghost. All these superfields are covariantly
(anti)chiral. Up to the second order in quantum superfields, the
ghost superfield action is given by
\be
S_{\rm gh}=\tr\int d^3x
d^4\theta\left[ -b(1+\Phi\square_-^{-1}\bar\Phi)\bar c +\bar
b(1+\bar\Phi\square_+^{-1}\Phi)c -\bar\varphi\varphi\right].
\label{gh-4}
\ee

The functional integral for the one-loop effective action reads
\be
e^{i\Gamma_{\cN=4}[{V},\Phi]}=e^{iS_{\cN=4}[{V},\Phi]}
\int {\cal D}v{\cal D}\phi {\cal D}b
{\cal D}c{\cal D}\varphi
e^{iS_2+iS_{\rm gf}+iS_{\rm gh}}\,.
\ee
Schematically, the one-loop effective action can be written as
\be
\Gamma_{\cN=4}=\frac i2{\rm Tr_v}\ln(\square_{\rm v}+\bar\Phi\Phi)
-i{\rm Tr}_+\ln (\square_++\bar\Phi\Phi)\,.
\label{N4EA}
\ee
The first term in the rhs in this expression comes from the
quantum gauge superfield while the second one takes into account
the contributions from quantum chiral superfield $\phi$ and ghosts.

\subsubsection{Gauge group SU$(N)$}
Now let us compute the traces of the logarithms of the operators
in (\ref{N4EA}) for the gauge group SU$(N)$ spontaneously broken
down to U$(1)^{N-1}$. The background gauge superfield $V$ is
specified in (\ref{V-back}). The background chiral superfield $\Phi$ has
similar structure,
\be
\Phi={\rm diag}({\bf\Phi}_1,{\bf\Phi}_2,\ldots,{\bf\Phi}_N)\,,\qquad
\sum_{I=1}^N {\bf\Phi}_I=0\,.
\label{phi-back}
\ee
The quantum gauge superfield $v$ is given by the expansion (\ref{basis})
while the quantum chiral superfield $\phi$ is represented by the expression similar
to (\ref{q-varphi}). Now it is straightforward to compute the
matrix traces in (\ref{N4EA}),
\be
\Gamma_{\cN=4}=i\sum_{I<J}^N{\rm Tr_v}\ln(
\hat\square_{{\rm v}\,IJ}+\bar{\bf\Phi}_{IJ}{\bf\Phi}_{IJ})
-2i\sum_{I<J}^N{\rm Tr}_+\ln (\hat\square_{+IJ}+\bar{\bf\Phi}_{IJ}
{\bf\Phi}_{IJ})\,,
\ee
where ${\bf\Phi}_{IJ}={\bf\Phi}_I-{\bf\Phi}_J$ and the operators
$\hat\square_{{\rm v}\,IJ}$ and $\hat\square_{+IJ}$ are given in
(\ref{v-square-ij}) and (\ref{square-ij}), respectively. The traces of the
logarithms of these operators are computed in a similar way as in
Sect.\ \ref{sect2.3}. As a result we get the one-loop effective
action in the $\cN=4$ SYM theory for the gauge group SU$(N)$ broken down
to U$(1)^{N-1}$,
\bea
\Gamma_{\cN=4}
&=&-\frac1\pi\sum_{I<J}^N\int d^3x d^4\theta\int_0^\infty\frac{ds}{s\sqrt{i\pi s}}
\frac{{\bf W}_{IJ}^2\bar{\bf W}_{IJ}^2}{{\bf B}_{IJ}^3}
e^{is({\bf G}_{IJ}^2+\bar{\bf \Phi}_{IJ}{\bf\Phi}_{IJ})}
\tanh\frac{s{\bf B}_{IJ}}{2}\sinh^2\frac{s{\bf B}_{IJ}}{2}\nn\\&&
-\frac{2}{\pi}\sum_{I<J}^N\int d^3x d^4\theta\bigg[
 {\bf G}_{IJ}\ln({\bf G}_{IJ}+\sqrt{{\bf G}_{IJ}^2+\bar{\bf \Phi}_{IJ}{\bf \Phi}_{IJ}})
  - \sqrt{{\bf G}_{IJ}^2+\bar{\bf \Phi}_{IJ}{\bf \Phi}_{IJ}}\nn\\&&
+\frac14\int_0^\infty\frac{ds}{\sqrt{i\pi s}}
e^{is({\bf G}_{IJ}^2+\bar{\bf \Phi}_{IJ}{\bf \Phi}_{IJ})}\frac{{\bf W}_{IJ}^2\bar{\bf W}_{IJ}^2}{{\bf B}_{IJ}^2}\left(
\frac{\tanh(s{\bf B}_{IJ}/2)}{s{\bf B}_{IJ}/2}-1
\right)\bigg].
\label{Gamma-N4}
\eea
We point out that only the leading terms given in the second line
in (\ref{Gamma-N4}) were studied in \cite{deBoer} by
employing the mirror symmetry while here we computed also all
higher order terms which are responsible in components for all higher powers
of the Maxwell field strength $F^{2n}$, $n\geq2$.

In conclusion of this section let us briefly comment on the case
of minimal gauge symmetry breaking ${\rm SU}(N)\to {\rm SU}(N-1)\times {\rm
U}(1)$. The background chiral superfield $\Phi$ is chosen
similarly as the gauge one (\ref{V-back1}),
\be
\Phi=\frac1{N}{\rm diag}\left((N-1){\bf \Phi},
\underbrace{-{\bf \Phi},\ldots,-{\bf \Phi}}_{N-1}\right).
\label{Phi-back2}
\ee
The leading terms in the $\cN=4$ SYM effective action in this case
are given by
\be
\Gamma_{\cN=4}
=\frac{2(N-1)}{\pi}\int d^3x d^4\theta\bigg[
  \sqrt{{\bf G}^2+\bar{\bf \Phi}{\bf \Phi}}
  - {\bf G}\ln({\bf G}+\sqrt{{\bf G}^2+\bar{\bf \Phi}{\bf \Phi}})
+\frac{1}{32}\frac{{\bf W}^2\bar{\bf W}^2}{({\bf G}^2+\bar{\bf \Phi}{\bf\Phi})^{5/4}}
+\ldots\bigg].
\label{3.21}
\ee
The first two terms in the rhs of this expression are responsible
for $\cN=4$ supersymmetric (and superconformal) generalization of the
Maxwell $F^2$ term
while the third term gives $F^4$ among other components
and the dots stand for higher-order terms.

Finally, let us comment on the following terms in the effective action
(\ref{3.21}),
\be
\int d^3x d^4\theta[
{\bf G}\ln({\bf G}+\sqrt{{\bf G}^2+\bar{\bf \Phi}{\bf \Phi}})
  - \sqrt{{\bf G}^2+\bar{\bf \Phi}{\bf \Phi}}]\,,
\label{dualGW}
\ee
which are known as the $\cN=2$, $d=3$ superspace action of the improved tensor
multiplet \cite{HKLR}. Note that analogous $\cN=1$, $d=4$ superspace action
of the improved tensor multiplet was constructed in \cite{LR}.
It is interesting to point out that (\ref{dualGW}) was obtained recently in \cite{KLL}
as a dual representation of the classical action of the Abelian
Gaiotto-Witten model \cite{GW} which is reviewed in the Appendix,
see (\ref{GW-dual}). Hence, the classical action of the Abelian
Gaiotto-Witten model in the representation (\ref{dualGW}) arises as the
leading term in the $\cN=4$ SYM effective action. Finally, we
point out that this term (\ref{dualGW}) in the $\cN=4$ SYM effective action
is known to be one-loop exact \cite{deBoer,deBoer1,Seib}.

\subsection{Low-energy effective action in $\cN=8$ SYM}
The classical $\cN=8$ SYM action appears by simple dimensional
reduction form the $\cN=4$, $d=4$ SYM. In our notations it reads
\be
S_{\cN=8}=\frac1{g^2}\tr\int d^3x d^4\theta \, [G^2
-\frac12 e^{-2V}\bar\Phi^i e^{2V} \Phi_i]
+\frac{1}{12g^2}\left(\tr\int d^3x d^2\theta \,\varepsilon^{ijk}
\Phi_i[\Phi_j,\Phi_k]+c.c.\right)
.
\label{N8}
\ee
Here $\Phi_i$, $i=1,2,3$, is a triplet of chiral superfields.
The action is invariant under the hidden $\cN=6$ supersymmetry with the
complex parameter $\epsilon_{\alpha\, i}$,
\bea
e^{-2V}\delta_\epsilon e^{2V}
&=&\theta^\alpha\epsilon_{\alpha\,i}\bar\Phi_c^i
-\bar\theta^\alpha\bar\epsilon_\alpha^i\Phi_{c\,i}\,,\nn\\
\delta_\epsilon\Phi_{c\,i}&=&-i\epsilon^\alpha_i \bar\nabla_\alpha G
+\frac14\varepsilon_{ijk}\bar\nabla^2(\bar\theta^\alpha\bar\epsilon_\alpha^j
\bar \Phi_c^k)
\,,\nn\\
\delta_\epsilon\bar\Phi_c^i&=&-i\bar\epsilon^{\alpha\,i} \nabla_\alpha G
+\frac14\varepsilon^{ijk}\nabla^2 (\theta^\alpha\epsilon_{\alpha\,j}\Phi_{c\,k})\,.
\label{6SUSY}
\eea
We use the notations $\bar\Phi_c^i=e^{-2V}\bar\Phi^i e^{2V}$,
$\Phi_c=\Phi$ for
the covariantly (anti)chiral superfields. Further we will omit the
subscript `$c$' assuming everywhere that we deal with covariantly
(anti)chiral superfields only.

The background field method can be easily generalized to the case
of the $\cN=8$ SYM theory. Let us sketch the basic steps.
The background-quantum splitting of the gauge superfield
(\ref{Vv}) is supplemented by the following splitting of the
(anti)chiral superfields,
\be
\Phi_i\to\Phi_i+g\phi_i\,,\qquad
\bar\Phi^i\to\bar\Phi^i+g\bar\phi^i\,,
\ee
with the corresponding `quantum' gauge transformations
\be
\delta \phi_i=i[\lambda,\frac1g\Phi_i+\phi_i]\,,\quad
\delta\bar\phi^i=i[\bar\lambda,\frac1g\bar\Phi^i+\bar\phi^i]\,,\quad
\delta\Phi_i=\delta\bar\Phi^i=0\,.
\label{q-tr2}
\ee
The gauge fixing functions are chosen in the form similar to
(\ref{f-f}),
\be
f= i\bar{\cal D}^2 v-\frac i2[\Phi_i,\bar{\cal
D}^2\square_-^{-1}\bar\phi^i]\,,\qquad
\bar f=i{\cal D}^2 v+\frac i2[\bar\Phi^i,{\cal D}^2\square_+^{-1}\phi_i]
\,.
\ee
When the background superfields are covariantly constant, $\bar{\cal
D}_\alpha\bar\Phi^i=0$, ${\cal D}_\alpha \Phi_i=0$, the quadratic
part of the action with respect to the quantum superfields takes
relatively simple form,
\bea
S_2+S_{\rm gf}&=&-\tr\int d^3x d^4\theta\left[
v(\square_{\rm v}+\bar\Phi^i\Phi_i)v+
\frac12\phi_i(\delta^i_j+\bar\Phi^i\Phi_j\square_-^{-1})\bar\phi^j
\right]\nn\\&&
+\frac14\left(
\tr\int d^3x d^2\theta\,\varepsilon^{ijk}\phi_i[\Phi_j,\phi_k] +c.c.
\right).
\eea
Here we denote $\bar\Phi^i\Phi_i v=[\bar\Phi^i[\Phi_i, v]]$. The
ghost superfield action is a simple generalization of
(\ref{gh-4}),
\be
S_{\rm gh}=\tr\int d^3x d^4\theta\left[
-b(1+\Phi_i\square_-^{-1}\bar\Phi^i)\bar c
+\bar b(1+\bar\Phi^i\square_+^{-1}\Phi_i)c
-\bar\varphi\varphi
\right].
\ee
As a result we see that the one-loop effective action is
relatively simple because it is defined by only one operator,
\be
\Gamma_{\cN=8}=\frac i2{\rm Tr_v}\ln(\square_{\rm v}+\bar\Phi^i\Phi_i)
\,.
\label{N8EA}
\ee
The contributions from ghosts and chiral superfields cancel each
other at one loop for the covariantly constant background
similarly as it happens for the $\cN=4$, $d=4$ SYM theory.

\subsubsection{Gauge group SU$(N)$}
For the gauge group SU$(N)$ spontaneously broken down to
U$(1)^{N-1}$ the gauge and chiral superfields are chosen
as in (\ref{V-back}) and (\ref{phi-back}). In this case the trace
of the logarithm in (\ref{N8EA}) is computed by standard methods
described in Sect.\ \ref{sect2.3},
\bea
\Gamma_{\cN=8}&=& i\sum_{I<I}^N{\rm Tr_v}
\ln(\hat\square_{{\rm v}\,IJ}+\bar{\bf\Phi}_{IJ}^i{\bf\Phi}_{i\,IJ})
\label{G8}\\
&=&-\frac1\pi\sum_{I<J}^N\int d^3x d^4\theta
\int_0^\infty\frac{ds}{s\sqrt{i\pi s}}
\frac{{\bf W}_{IJ}^2\bar{\bf W}_{IJ}^2}{{\bf B}_{IJ}^3}e^{is({\bf G}_{IJ}^2
+\bar{\bf\Phi}_{IJ}^i{\bf\Phi}_{i\,IJ})}
\tanh\frac{s{\bf B}_{IJ}}{2}\sinh^2\frac{s{\bf B}_{IJ}}{2}\,.\nn
\eea

In the case when the gauge group SU$(N)$ is spontaneously broken
down to SU$(N-1)\times$U$(1)$, the background superfields should
be chosen as in (\ref{V-back1}) and (\ref{Phi-back2}). Then the
leading term in the effective action (\ref{G8}) is given by
\be
\Gamma_{\cN=8}=\frac{3(N-1)}{32\pi}\int d^3x d^4\theta
\frac{{\bf W}^2\bar{\bf W}^2}{({\bf G}^2+\bar{\bf\Phi}^i{\bf\Phi}_i)^{5/2}}
+\ldots\sim \int d^3x\frac{(F^{mn}F_{mn})^2}{(f^{\ui} f_{\ui})^{5/2}}+\ldots\,,
\label{G8lead}
\ee
where $f^{\ui}$, $\ui=1,2,\ldots,7$ are the seven scalar fields in the
$\cN=8$ SYM theory and dots stand for the higher-order terms.
In \cite{DS} it was argued that the $F^4$ term in the $\cN=8$ SYM
effective action (\ref{G8lead}) is one-loop exact in the
perturbation theory, but it receives instanton corrections.

Since there is no $F^2$ term in (\ref{G8lead}),
the ABJM theory in the dual representation
(\ref{ABJM-dual-1}) cannot be a part of the low-energy effective
action of the $\cN=8$ SYM theory. This is quite expected because
the $\cN=8$ SYM describes the D2 brane in the flat background while
the Abelian ABJM theory corresponds to the M2 brane on a
Calabi-Yau fourfold $X_4={\mathbb C}^4/{\mathbb Z}_k$ \cite{ABJM}.
Moreover, the ABJM theory has only $\cN=6$ supersymmetry while the
effective action (\ref{G8lead}) should respect the $\cN=8$
supersymmetry since the supersymmetry cannot be broken within the
quantization.
In the next section we will see that the ABJM theory does appear in
the low-energy effective action when one considers
the D2 brane in the appropriate background.

\setcounter{equation}{0}
\section{Low-energy effective action in $\cN=2$ quiver gauge theories}
One of the ways to understand multiple 2-branes in M theory is
through studying the strong coupling limit of a stack of D2 branes
in IIA string theory. However, one has to specify the backgrounds in
which the M2 and D2 branes live. Indeed, in the previous section we
have shown that the $\cN=8$ SYM theory cannot reproduce the ABJM theory
because the latter describes the M2 brans on the conifold
$X_4={\mathbb C}^4/{\mathbb Z}_k$ while the former corresponds to D2
branes in flat space. In \cite{Aganagic} it was explained that the
ABJM theory should appear in the strong coupling limit of a stack
of D2 branes probing the singularity of the Calabi-Yau threefold
$X_3$ fibred over real line $\mathbb R$ and with RR 2-form fluxes
turned on. Here $X_3$ (or $Y_6$ in the notation of \cite{KW}) is a
cone with the base $T^{1,1}=({\rm SU}(2)\times{\rm SU}(2))/{\rm
U}(1)$ \cite{DNP}. This conifold $X_3$ fibred over $\mathbb R$
appears as the moduli space of the field theory describing these D2
branes. The field theory in question is a three-dimensional quiver
gauge theory given by the $\cN=2$ SYM with gauge group
SU$(N)\times$SU$(N)$ and with the matter given by four chiral
superfields in the $(N,\bar N)$ and $(\bar N,N)$
bifundamental representations of the gauge group.

In fact, the interest to the three-dimensional quiver gauge
models was initiated by the works \cite{KW,Kachru} where
similar four-dimensional supergauge theory was studied in details as
a model for multiple D3 branes probing the singularity of $X_3$
(see, e.g., \cite{KT} for a recent review), but the considerations of D2 branes
on this background go along similar lines. It is useful to give
here a short review of the moduli space for such three-dimensional field
theories because it
motivates the choice of the background superfields which we do in
the following studies of the effective action. To understand the
moduli space it is sufficient to take the Abelian version of the theory,
i.e., when the gauge group is U$(1)_L\times$U$(1)_R$. The matter
superfields are given by a pair of chiral superfields
(further referred to as the hypermultiplets\footnote{To be precise,
the pair of (anti)chiral
superfields $(\bar Q_+,Q_-)$ form a hypermultiplet only if one
considers the $\cN=4$ supersymmetric gauge theory. Here we have only
$\cN=2$ supersymmetry because the gauge superfields do not get their
$\cN=4$ partners, but we hope that our terminology will not be
misleading.}) $(\bar Q_+^a,Q_-^a)$ which are labeled by the SU(2)
index $a=1,2$. The classical action corresponds to the massless
$\cN=2$ electrodynamics, \be S=\frac1{g^2}\int d^3x
d^4\theta(G_L^2+G_R^2) -\frac12\int d^3x d^4\theta(\bar Q_+^a
e^{2(V_L-V_R)}Q_{+a} +Q_-^a e^{2(V_R-V_L)}\bar Q_{-a})\,,
\label{S-ab} \ee where $V_L$ and $V_R$ are the `left' and `right'
Abelian gauge superfields with superfield strengths $G_L$ and $G_R$
respectively. Among other components, the gauge superfields contain
the real scalars $\phi_{L,R}$ and auxiliary fields $D_{L,R}$,
\be
V_{L,R}=i\bar\theta^\alpha\theta_\alpha\phi_{L,R}+\bar\theta^2\theta^2
D_{L,R}+\ldots,
\ee
while the scalars in the chiral superfields are
\be
\bar Q_+^a=\bar z_+^a+\ldots\,,\qquad
Q_-^a= z_-^a +\ldots\,.
\ee
Eliminating the auxiliary fields $D_{L,R}$ one can see that the
potential for the scalars has minimum when two conditions are
satisfied,
\be
(i)\quad | z_{+1}|^2+|
z_{+2}|^2=|z_-^1|^2+|z_-^2|^2\,,\qquad
(ii)\quad \phi_L=\phi_R\,.
\label{iii}
\ee
Taking into account the gauge transformations for
the scalars, the identification
\be
z_{+a}\sim
z_{+a}e^{i\alpha}\,,\qquad z_{-a}\sim z_{-a}e^{-i\alpha}\,,
\label{fff}
\ee
considered together with the first constraint in
(\ref{iii}) defines precisely the conifold denoted by $X_3$ (see
\cite{KW}), while the second constraint in (\ref{iii}) is just the
real line. Therefore the moduli space of the model (\ref{S-ab}) is
$X_3\times \mathbb R$. The non-Abelian version of this model has
the same moduli space modulo the action of the Weyl group. Note that
in the non-Abelian case the SU(2)-invariant superpotential should be
added to the action (\ref{S-ab}), but it does not change the moduli.

In what follows, we will study the low-energy effective actions in
two non-Abelian $\cN=2$ quiver gauge theories, namely with one and with
two hypermultiplets. The latter model corresponds to the D2 branes
probing the conifold $X_3\times \mathbb R$ while the former is
simple and helps us to understand the basic steps of the
quantization procedure. The resulting effective actions will depend
on the background Abelian superfields satisfying the superfield
analogs of the constraints (\ref{iii}).
We point out that effective action is gauge independent
when the background superfields correspond to the vacuum of the
model.

\subsection{$\cN=2$ SYM with one bifundamental hypermultiplet}
One of the simplest examples of quiver gauge theories is the
$\cN=2$ SYM theory with twisted gauge group ${\cal G}_L\times {\cal
G}_R$ and with a hypermultiplet $(\bar Q_+,Q_-)$ in the
bifundamental representation of this gauge group. The classical
action reads
\be
S=\frac{1}{g^2}\tr\int d^3x d^4\theta(G_L^2+G_R^2)
-\frac1{2}\tr\int d^3x d^4\theta (\bar Q_+ e^{2V_L} Q_+ e^{-2V_R}  +Q_- e^{-2V_L}
 \bar Q_- e^{2V_R} )\,.
\label{N2-bifund}
\ee
Here $V_L$ and $V_R$ are the gauge superfields taking theirs
values in the Lie algebras of the `left' ${\cal G}_L$ and `right' ${\cal
G}_R$ gauge groups, respectively. For simplicity, we consider the kinetic terms
for these gauge superfields with the same gauge coupling $g$ and
do not give any superpotential for the chiral superfields.
The classical action (\ref{N2-bifund}) is invariant under the following gauge
transformations with the (anti)chiral gauge parameters $\lambda_{L,R}$,
$\bar\lambda_{L,R}$,
\bea
e^{2V_L}\to e^{i\bar\lambda_L}e^{2V_L}e^{-i\lambda_L}\,,&\quad&
e^{2V_R}\to e^{i\bar\lambda_R}e^{2V_R}e^{-i\lambda_R}\,,\nn\\
Q_+\to e^{ i\lambda_L} Q_+ e^{-i\lambda_R}\,,&\quad&
Q_-\to e^{ i\lambda_R}Q_- e^{-i\lambda_L}\,.
\label{4.2}
\eea
The low-energy effective action in analogous four-dimensional
theory was studied in particular in \cite{Kuz}.

To derive the structure of one-loop effective action we perform
the background-quantum splitting of the gauge and matter
superfields,
\be
e^{2V_L}\to e^{2 V_L}e^{2gv_L}\,,\quad
e^{2V_R}\to e^{2 V_R}e^{2gv_R}\,,\quad
Q_\pm \to Q_\pm +q_\pm\,,
\ee
where $V_{L,R}$, $Q_{\pm}$ on the right denote the background superfields and
$v_{L,R}$, $q_{\pm}$ are the quantum ones. The gauge
transformations for the quantum gauge superfields have the form
(\ref{q-tr}). However, now we need two gauge fixing functions to break the
gauge invariance under the left and the right gauge groups,
\be
f_L=i\bar{\cal D}^2v_L\,,\qquad
f_R=i\bar{\cal D}^2v_R\,.
\ee
The corresponding gauge fixing action reads (we use the Fermi-Feynman gauge)
\be
S_{\rm gf}=\frac18\tr\int d^3x d^4\theta(\bar f_L f_L+\bar f_R
f_R)=-\frac1{16}\tr\int d^3x d^4\theta
(v_L\{{\cal D}^2,\bar{\cal D}^2 \}v_L+
v_R\{{\cal D}^2,\bar{\cal D}^2 \}v_R)\,.
\ee
The background gauge covariant spinor derivatives ${\cal D}_\alpha$, $\bar{\cal
D}_\alpha$ act appropriately on the left and right gauge
superfields and chiral superfields, e.g.,
\bea
&{\cal D}_\alpha v_L=D_\alpha v_L+[V_{L\alpha},v_L]\,,\qquad
{\cal D}_\alpha v_R=D_\alpha v_R+[V_{R\alpha},v_R]\,,&\nn\\
&{\cal D}_\alpha Q_+=D_\alpha Q_++ V_{L\alpha} Q_+ -Q_+
V_{R\alpha}\,.&
\label{4.11}
\eea

Expanding the action (\ref{N2-bifund}) up to the second order in
the quantum superfields, we get the action $S_2$ which, together
with $S_{\rm gf}$, is
\bea
S_2+S_{\rm gf}&=&-\tr\int d^3x d^4\theta\big[ v_L(\square_{\rm v}+g^2 Q_+\bar Q_+
+g^2 \bar Q_- Q_-)v_L\nn\\&&
+v_R(\square_{\rm v}+g^2\bar Q_+ Q_+
+ g^2 Q_-\bar Q_-)v_R-2g^2\bar Q_+ v_L Q_+ v_R -2g^2 Q_- v_L\bar Q_- v_R
\big]\nn\\
&&
-\tr\int d^3x d^4\theta\big[ \frac12\bar q_+ q_+ +\frac12q_-\bar q_- \nn\\&&
 +g(\bar q_+ v_L Q_+-\bar q_+ Q_+ v_R-\bar Q_+ q_+ v_R
 +\bar Q_+ v_L q_+)\nn\\&&
 +g(Q_-\bar q_- v_R - Q_- v_L\bar q_- - q_- v_L\bar Q_-
 +q_- \bar Q_- v_R )
\big]\,.
\label{S2-bifund}
\eea
Note that we use the superfields $\bar Q_\pm$ and $\bar q_\pm$ in this action
which are covariantly antichiral, ${\cal D}_\alpha\bar Q_\pm={\cal D}_\alpha \bar
q_\pm=0$, where the covariant spinor derivative acts in accord
with the bifundamental representation of the gauge group (\ref{4.11}).

The action (\ref{S2-bifund}) defines the structure of the one-loop effective action
\be
e^{i\Gamma}=\int {\cal D}v_{L,R}{\cal D}q_\pm
{\cal D}b_{L,R}{\cal D}c_{L,R}{\cal D}\varphi_{L,R}
e^{i(S_2+S_{\rm gf}+S_{\rm gh})}\,,
\label{Gamma-bifund}
\ee
where $S_{\rm gh}$ is the quadratic part of the ghost superfield
action with respect to the quantum superfields,
\be
S_{\rm gh}=\tr\int d^3x d^4\theta [ \bar b_L c_L- b_L\bar c_L
+\bar b_R c_R- b_R\bar c_R+\bar \varphi_L\varphi_L
+\bar\varphi_R\varphi_R]\,.
\label{bi-ghosts}
\ee
The superfields $b_{L,R}$, $c_{L,R}$ are the Faddeev-Popov ghosts
while $\varphi_{L,R}$ are the Nielsen-Kallosh ghosts. All these ghost
superfields are covariantly (anti)chiral with respect to either left or
right gauge group.

The generating functional (\ref{Gamma-bifund}) gives a general
expression for the one-loop effective action. To get the concrete
result we have to fix the gauge groups ${\cal G}_L$ and ${\cal G}_R$ and to
specify the background superfields.

\subsubsection{Gauge group SU$(2)\times$SU$(2)$}
\label{441}
This is the simplest non-Abelian gauge group. The superfields
$v_{L}$ and $v_{R}$ take their values in the Lie algebras of the
gauge groups SU$(2)_L$ and SU$(2)_R$, respectively, while the chiral
superfields are general complex $2\times 2$ matrices. We choose
the background gauge superfields to belong to the Cartan
subalgebras,
\be
V_L=V_R =\frac12\left(
\begin{array}{cc}
{\bf V}&0\\0&-{\bf V}
\end{array}
\right).
\label{V-bg}
\ee
Here $\bf V$ is the usual Abelian gauge superfield with the
superfield strength $\bf G$.
The background chiral superfields are chosen as follows
\be
Q_+=
\left(
\begin{array}{cc}
{\bf Q}_+ & 0\\0&0
\end{array}
\right),\qquad
Q_-=
\left(
\begin{array}{cc}
{\bf Q}_- & 0\\0&0
\end{array}
\right).
\label{Q-bg}
\ee
Moreover, the superfields ${\bf Q}_+$ and ${\bf
Q}_-$ obey the following constraint
\be
\bar{\bf Q}_+{\bf Q}_+ = \bar{\bf Q}_- {\bf Q}_-
\label{Q-constr}
\,,
\ee
which is a superfield analog of (\ref{iii}).
Note that the matrices (\ref{V-bg}) and (\ref{Q-bg})
commute. Hence, the constraint of covariant antichirality of the
superfields $\bar{\bf Q}_\pm$ turns into the usual
antichirality, $D_\alpha \bar{\bf Q}_\pm=0$.

Now we specify the constraints on the background gauge $\bf V$ and
matter ${\bf Q}_\pm$ superfields. We will be interested in the leading part of
the effective action $\Gamma$ which depends on the superfield
strength $\bf G$ and (anti)chiral superfields $\bar{\bf Q}_\pm$,
${\bf Q}_\pm$, but not on their derivatives,
\be
\Gamma=\Gamma[{\bf G},\bar{\bf Q}_\pm,
{\bf Q}_\pm]=\int d^3x d^4\theta\,{\cal L}_{\rm eff}({\bf G},\bar{\bf Q}_\pm,
{\bf Q}_\pm)\,.
\ee
Note that we neglect not only the space-time derivatives of the
superfields, but also all their covariant spinor derivatives,
\be
D_\alpha {\bf Q}_\pm=0\,,\quad
\bar D_\alpha \bar{\bf Q}_\pm=0\,,\quad
D_\alpha{\bf G}=\bar {\bf W}_\alpha=0\,,\quad
\bar D_\alpha{\bf G}={\bf W}_\alpha=0\,.
\label{Q-constr1}
\ee
This constraint is sufficient to study the contribution to the
effective action of the second order with respect to the Maxwell
field strength, $F^2$. However, two last constraints in
(\ref{Q-constr1}) should be taken carefully. Indeed, one can
freely neglect the superfield strengths ${\bf W}_\alpha$, $\bar{\bf
W}_\alpha$ while doing the computations in full $\cN=2$
superspace, but this is not true in the chiral superspace. To be precise,
one has to keep these superfield strengths in computing the trace
of logarithm of the chiral box operator $\hat\square_+$ and to vanish
them only after passing to the full superspace measure. We will
keep this in mind while doing the corresponding computations with
the chiral box operator.

To proceed, we have to do the matrix traces in the action
(\ref{S2-bifund}) for the chosen background matrix superfields
(\ref{V-bg}) and (\ref{Q-bg}). The quantum superfields
$v_{L,R}$ and $q_{\pm}$ have both diagonal and off-diagonal matrix
components,
\be
v_{L,R}=v_{L,R({\rm diag})}+v_{L,R({\rm off-diag})}\,,\qquad
q_\pm=q_{\pm({\rm diag})}+q_{\pm({\rm off-diag})}\,.
\ee
It is an easy exercise to check that the action (\ref{S2-bifund})
appears to be a sum of two independent parts,
\be
S_2[v_{L,R},q_\pm]+S_{\rm gf}[v_{L,R},q_\pm]
=S_{\rm diag}[v_{L,R({\rm diag})},q_{\pm({\rm diag})}]
+S_{\rm off-diag}[v_{L,R({\rm off-diag})},q_{\pm({\rm off-diag})}]\,.
\label{4.21}
\ee
Therefore the contributions to the one-loop effective action from
the diagonal matrix components of the quantum superfields can be studied
independently of the off-diagonal ones. Further we will write down
the expressions for the actions $S_{\rm diag}$ and $S_{\rm
off-diag}$ explicitly and compute the corresponding contributions
to the effective action.

Consider first only the diagonal components,
\be
v_{L,R}=\frac12\left(
\begin{array}{cc}
{\rm v}_{L,R} & 0 \\ 0& -{\rm v}_{L,R}
\end{array}
\right),\qquad
q_{\pm}=\left(
\begin{array}{cc}
{\rm q}_{\pm} & 0 \\ 0& {\rm p}_{\pm}
\end{array}
\right),
\label{vq-diag}
\ee
where ${\rm v}_{L,R}$ are real while ${\rm p}_\pm$ and ${\rm
q}_\pm$ are chiral, $\bar D_\alpha{\rm
q}_\pm=\bar D_\alpha{\rm
p}_\pm=0$.
The part of the action (\ref{S2-bifund}) for these superfields reads
\bea
S_{\rm diag}&=&-\frac12\int d^3x d^4\theta\big[
{\rm v}_L(\square+M^2){\rm v}_L+ {\rm v}_R(\square+M^2){\rm v}_R
-2{\rm v}_L{\rm v}_R M^2\nn\\&&
+\bar{\rm q}_{+}{\rm q}_{+}+\bar{\rm p}_{+}{\rm p}_{+}
+\bar{\rm q}_{-}{\rm q}_{-}+\bar{\rm p}_{-}{\rm p}_{-}\nn\\&&
+g({\rm v}_L-{\rm v}_R)( {\bf Q}_+\bar {\rm q}_{+} +\bar {\bf Q}_+{\rm q}_{+}
-{\bf Q}_-\bar{\rm q}_{-} -\bar{\bf Q}_-{\rm q}_{-})
\big]\,,
\label{S-diag}
\eea
where
\be
M^2=g^2 \bar{\bf Q}_+{\bf Q}_+=g^2 \bar{\bf Q}_-{\bf Q}_-
\label{M2}
\ee
is the effective mass squared of the gauge superfields. Note that
the superfields ${\rm p}_{\pm}$ are completely free and do not
contribute to the effective action.

Despite the mixing of the gauge and matter superfields in
(\ref{S-diag}), it is quite straightforward to diagonalize the matrix
of second variational derivatives of this action with respect
to the quantum superfields. The details of these computations are
given in the Appendix \ref{ApB1}. Here we present only the result,
\be
\Gamma_{\rm diag}=
\frac i2 {\rm Tr_v}\ln\left(1-\frac 14\frac{M^2}{\square^2}D^\alpha\bar D^2 D_\alpha\right)
=-\frac i{16}{\rm Tr_v}\frac1\square\ln\left(1+\frac{2M^2}{\square}\right)D^\alpha
\bar D^2 D_\alpha\,.
\label{4.25}
\ee
This is a functional trace of the operator acting in the full
superspace which can be easily computed using the following
relations
\bea
D^\alpha \bar D^2 D_\alpha \delta^7(z-z')|_{\theta=\theta'}
&=&16\delta^3(x-x')\,,\nn\\
\frac1\square\ln(1+M^2/\square)\delta^3(x-x')|_{x=x'}&=&\frac i{2\pi}
\sqrt{M^2}\,.
\label{4.26}
\eea
As a result we get
\be
\Gamma_{\rm diag}=\frac g{2\pi}\int d^3x d^4\theta\sqrt{\bar{\bf Q}_+{\bf Q}_+
+\bar{\bf Q}_-{\bf Q}_-}\,.
\label{G-diag}
\ee
This is nothing but the effective K\"ahler potential which
coincides up to a coefficient with the one in the three-dimensional $\cN=2$
supersymmetric electrodynamics \cite{AHKSS}. Note that the
constraint (\ref{Q-constr}) is not necessary for this part of the
effective action and even if one relaxes this, the result will be
the same. However, this constraint will be necessary for computing
the part of the effective action for the off-diagonal superfields
in an unambiguous way.

Let us now concentrate on the contributions to the effective
action from the off-diagonal superfields,
\be
v_{L,R}=\frac12\left(
\begin{array}{cc}
0 & {\rm v}_{L,R}  \\ \bar{\rm v}_{L,R} &0
\end{array}
\right),\qquad
q_{\pm}=\left(
\begin{array}{cc}
0 & {\rm q}_{\pm} \\ {\rm p}_{\pm} & 0
\end{array}
\right).
\label{4.27}
\ee
Here we use the same letters ${\rm v}_{L,R}$ and ${\rm q}_\pm$, ${\rm p}_{\pm}$ as for the diagonal
superfields in (\ref{vq-diag}) despite they are completely
independent. We hope that this will not lead to any confusions
because there are no mixing between the diagonal and off-diagonal
components and the corresponding parts of the classical action
can be studied independently.
The part of the action (\ref{S2-bifund}) for these superfields
reads
\bea
S_{\rm off-diag}&=&-\frac12\int d^3x d^4\theta\big[
\bar{\rm v}_L(\hat\square_{\rm v}+M^2){\rm v}_L
+\bar{\rm v}_R(\hat\square_{\rm v}+M^2){\rm v}_R\nn\\&&
+\bar{\rm q}_+{\rm q}_+ +\bar{\rm q}_-{\rm q}_-
+\bar{\rm p}_+{\rm p}_+ +\bar{\rm p}_-{\rm p}_-\nn\\&&
+g{\bf Q}_+(\bar{\rm p}_+ \bar{\rm v}_L-\bar{\rm q}_+{\rm v}_R)
+g\bar{\bf Q}_+({\rm p}_+{\rm v}_L-{\rm q}_+\bar{\rm v}_R)\nn\\&&
+g{\bf Q}_-(\bar{\rm p}_-\bar{\rm v}_R-\bar{\rm q}_- {\rm v}_L)
+g\bar{\bf Q}_-({\rm p}_-{\rm v}_R-{\rm q}_-\bar{\rm v}_L)
\big]\,.
\label{S-off}
\eea
Note that the superfields $\bar{\rm p}_\pm $ and $\bar{\rm q}_\pm$
are covariantly antichiral, ${\cal D}_\alpha \bar{\rm p}_\pm
={\cal D}_\alpha \bar{\rm q}_\pm=0$, where the gauge covariant
spinor derivative is defined with the proper charge for
the ``$+$'' and ``$-$'' superfields. The operator $\hat\square_{\rm
v}$ has the form (\ref{v-square-ij}), but without indices $IJ$
since there is only one component $\bf V$ in the background gauge
superfield $V$ in the case of the gauge group SU(2).

Now it is straightforward to compute the trace of the logarithm of
the matrix of second variational derivatives for the action
(\ref{S-off}). The details of this procedure are collected in the
Appendix \ref{ApB2}, the net result is
\be
\Gamma_{\rm off-diag}=2i{\rm Tr}_+\ln\hat\square_++
2i{\rm Tr_v}\ln\hat\square_{\rm v}
+2i{\rm Tr_v}\ln\left(1-\frac{M^2}{8\hat\square_{\rm v}^2}{\cal D}^\alpha\bar{\cal D}^2
{\cal D}_\alpha\right)\,.
\label{4.30}
\ee
The trace of the logarithm of the operator $\hat\square_{\rm v}$
was studied in Sect.\ \ref{sect2.3}. The corresponding
contributions to the effective action start from $F^4$ and
therefore they are irrelevant for our consideration. Next, we have the
contribution
\be
2i{\rm Tr_v}\ln\left(1-\frac{M^2}{8\hat\square_{\rm v}^2}{\cal D}^\alpha\bar{\cal D}^2
{\cal D}_\alpha\right)
=-4i\Tr\frac1{\square+{\bf G}^2}\ln(1+\frac{M^2}{\square+{\bf G}^2})\,,
\label{4.31}
\ee
which is derived in (\ref{B15}). The functional trace in
(\ref{4.31}) results in the following momentum integral
\be
\int_0^\infty\frac{k^2 dk}{k^2+{\bf G}^2}\ln(1+\frac{M^2}{k^2+{\bf G}^2})
=\pi\left[{\bf G}\ln{\bf G}+\sqrt{{\bf G}^2+M^2}-{\bf G}\ln({\bf G}
+\sqrt{{\bf G}^2+M^2})\right]\,.
\ee
The expression (\ref{4.30}) contains also the trace of the
logarithm of the chiral box operator which was
studied in details in \cite{BPS}.  Up to a coefficient, it
is given by (\ref{Gh}).
Summing up these contributions, we get the part of the effective
action induced by the off-diagonal superfields,
\be
\Gamma_{\rm off-diag}
=\frac 1\pi\int d^3x d^4\theta[
3{\bf G}\ln{\bf G}+2\sqrt{{\bf G}^2+g^2\bar{\bf Q}_+{\bf Q}_+}-2{\bf G}\ln({\bf G}
+\sqrt{{\bf G}^2+g^2\bar{\bf Q}_+{\bf Q}_+})]\,.
\label{G-off}
\ee

Finally, we have to take into account the contributions from the
ghost superfields (\ref{bi-ghosts}). This is exactly the result
(\ref{Gh}) considered for $n=2$ because the gauge group is SU(2)
and doubled because we have the left-right pairs of the ghosts,
\be
\Gamma_{\rm ghosts}=-\frac3\pi\int d^3x d^4\theta\,{\bf G}\ln{\bf G}\,.
\label{G-gh}
\ee

Summing up (\ref{G-diag}), (\ref{G-off}) and (\ref{G-gh})
together, we get the low-energy effective action in the model
(\ref{N2-bifund}),
\bea
\Gamma&=&\frac g{2\pi}\int d^3x d^4\theta\sqrt{\bar{\bf Q}_+{\bf Q}_+
+\bar{\bf Q}_-{\bf Q}_-}\nn\\&&
+\frac2\pi \int d^3x d^4\theta\left[
\sqrt{{\bf G}^2+g^2\bar{\bf Q}_+{\bf Q}_+}-{\bf G}\ln({\bf G}+\sqrt{{\bf G}^2+g^2\bar{\bf Q}_+{\bf
Q}_+})\right].
\label{G}
\eea
The terms in the last line of (\ref{G}) have the form
(\ref{GW-dual}) which is the dualized
classical action in the Abelian Gaiotto-Witten model. Therefore
they possess the $\cN=4$ supersymmetry of the Gaiotto-Witten
model. However, they are corrected by the K\"ahler potential
appearing in the first line in (\ref{G}) which has only the
$\cN=2$ supersymmetry.

\subsection{$\cN=2$ SYM with two bifundamental hypermultiplets}
Let us consider the generalization of the model (\ref{N2-bifund})
to the case when the hypermultiplet has the SU(2) doublet index,
$(\bar Q_+^a, Q_{-}^a)$, $a=1,2$, and the action (\ref{N2-bifund})
is extended with a SU(2) invariant superpotential,
\bea
S&=&\frac{1}{g^2}\tr\int d^3x d^4\theta(G_L^2+G_R^2)
-\frac1{2}\tr\int d^3x d^4\theta (\bar Q_+^a e^{2V_L} Q_{+a} e^{-2V_R}
+Q_{-}^a e^{-2V_L} \bar Q_{-a} e^{2V_R} )
\nn\\&&
-\frac\lambda2\left(\tr\int d^3x
d^2\theta\,\varepsilon^{ab}\varepsilon_{cd}Q_{+a}Q_{-}^cQ_{+b}Q_{-}^d
+c.c.
\right).
\label{N2-bifund2}
\eea
Here $\lambda$ is the dimensionless coupling constant.
The gauge symmetry in this case is similar to
(\ref{4.2}) with obvious insertion of the indices for the
hypermultiplet. Therefore, the gauge fixing and structure of
ghosts is the same as for the model (\ref{N2-bifund}). Doing the
background-quantum splitting in the standard way, $e^{2V_{L,R}}\to e^{2V_{L,R}}
e^{2gv_{L,R}}$,
$Q_{\pm a}\to Q_{\pm a}+q_{\pm a}$, we get the following quadratic
action for the quantum superfields,
\bea
S_2+S_{\rm gf}&=&S_{\rm gauge}+S_{\rm hyper}+S_{\rm pot}\,,\label{S2-bifund-2}
\\
S_{\rm gauge}&=&-\tr\int d^3x d^4\theta\big[ v_L(\square_{\rm v}+g^2 Q_{+a}\bar
Q_+^a+g^2 \bar Q_{-a} Q_-^a)v_L\nn\\&&
+v_R(\square_{\rm v}+g^2\bar Q_+^a Q_{+a}
+ g^2 Q_-^a\bar Q_{-a})v_R
-2g^2\bar Q_+^a v_L Q_{+a} v_R -2g^2 Q_-^a v_L\bar Q_{-a} v_R
\big]\,,\nn\\
S_{\rm hyper}&=&
-\tr\int d^3x d^4\theta\big[ \frac12\bar q_+^a q_{+a} +\frac12q_-^a\bar q_{-a} \nn\\&&
 +g(\bar q_+^a v_L Q_{+a}-\bar q_+^a Q_{+a} v_R-\bar Q_+^a q_{+a} v_R
 +\bar Q_+^a v_L q_{+a})\nn\\&&
 +g(Q_-^a\bar q_{-a} v_R - Q_-^a v_L\bar q_{-a} - q_-^a v_L\bar
 Q_{-a}  +q_-^a \bar Q_{-a} v_R )
\big]\,,
\nn\\
S_{\rm pot}&=&
-\frac\lambda2\tr\int d^3x d^2\theta\,\varepsilon^{ab}\varepsilon_{cd}
[2Q_{+a}Q_{-}^c q_{+b}q_{-}^d+
2Q_{+a}q_{-}^c q_{+b}Q_{-}^d\nn\\&&
+Q_{+a}q_{-}^c Q_{+b}q_{-}^d+q_{+a}Q_{-}^c
q_{+b}Q_{-}^d]+c.c.\nn
\eea
This action defines the structure of the one-loop effective action
for the background superfields. To get the concrete result we have
to specify the gauge group and the background.

\subsubsection{Gauge group SU$(2)\times$SU$(2)$}
The background gauge superfields for this gauge group are chosen
in the form (\ref{V-bg}) while for the background matter
superfields we take
\be
Q_{+a}=
\left(
\begin{array}{cc}
{\bf Q}_{+a} & 0\\0&0
\end{array}
\right),\qquad
Q_-^a=
\left(
\begin{array}{cc}
{\bf Q}_-^a & 0\\0&0
\end{array}
\right),
\label{Q-bg-2}
\ee
where ${\bf Q}_\pm$ are chiral superfields,
$\bar D_\alpha {\bf Q}_{+a}=\bar D_\alpha {\bf Q}_-^a=0$.
The constraint (\ref{Q-constr}) in this case turns into
\be
\bar{\bf Q}_+^a{\bf Q}_{+a} = \bar{\bf Q}_{-a} {\bf Q}_-^a
\label{Q-constr-2}
\,.
\ee
The background superfields obey also the
constraint (\ref{Q-constr1}) because we are interested in the
contributions to the effective action of the order $F^2$.

A nice feature of the background (\ref{V-bg},\ref{Q-bg-2}) is that
the diagonal matrix components of the quantum gauge and matter
superfields completely decouple from the off-diagonal ones
so that the relation (\ref{4.21}) holds for the action (\ref{S2-bifund-2}) as
well. Similarly as for the model (\ref{S2-bifund}), we study first the
contributions to the effective action from the diagonal
matrix components,
\be
v_{L,R}=\frac12\left(
\begin{array}{cc}
{\rm v}_{L,R} & 0 \\ 0& -{\rm v}_{L,R}
\end{array}
\right),\quad
q_{+a}=\left(
\begin{array}{cc}
{\rm q}_{+a} & 0 \\ 0& {\rm p}_{+a}
\end{array}
\right),\quad
q_{-}^a=\left(
\begin{array}{cc}
{\rm q}_{-}^a & 0 \\ 0& {\rm p}_{-}^a
\end{array}
\right),
\label{vq-diag-2}
\ee
where ${\rm q}_\pm$ and ${\rm p}_\pm$ are standard chiral superfields,
$\bar{D}_\alpha{\rm q}_\pm=\bar D_\alpha{\rm p}_\pm=0$.
The part of the classical action for these superfields reads
\bea
S_{\rm diag}&=&-\frac12\int d^3x d^4\theta\big[
{\rm v}_L(\square+M^2){\rm v}_L+ {\rm v}_R(\square+M^2){\rm v}_R
-2{\rm v}_L {\rm v}_R M^2\nn\\&&
+\bar{\rm q}_{+}^a{\rm q}_{+a}+\bar{\rm p}_{+}^a{\rm p}_{+a}
+\bar{\rm q}_{-a}{\rm q}_{-}^a+\bar{\rm p}_{-a}{\rm p}_{-}^a\nn\\&&
+g({\rm v}_L-{\rm v}_R)( {\bf Q}_{+a}\bar {\rm q}_{+}^a +\bar {\bf Q}_+^a{\rm q}_{+a}
-{\bf Q}_-^a\bar{\rm q}_{-a} -\bar{\bf Q}_{-a}{\rm q}_{-}^a)
\big]\,,
\label{S-diag-2}
\eea
where
\be
M^2=g^2 \bar{\bf Q}_+^a{\bf Q}_{+a}=g^2 \bar{\bf Q}_{-a}{\bf
Q}_-^a\,.
\label{4.37}
\ee
The action (\ref{S-diag-2}) is very similar to (\ref{S-diag}).
Therefore we can easily generalize the K\"ahler potential
(\ref{G-diag}) to the case of the model with two hypermultiplets,
\be
\Gamma_{\rm diag}=
\frac i2 {\rm Tr_v}\ln\left(1-\frac{M^2}{4\square^2}D^\alpha\bar D^2 D_\alpha \right)
=\frac{\sqrt 2 g}{2\pi}\int d^3x d^4\theta\sqrt{\bar{\bf Q}_{+a}{\bf Q}_+^a}\,.
\label{G-diag-2}
\ee

Now let us consider the off-diagonal quantum superfields,
\be
v_{L,R}=\frac12\left(
\begin{array}{cc}
0 & {\rm v}_{L,R}  \\ \bar{\rm v}_{L,R} &0
\end{array}
\right),\quad
q_{+a}=\left(
\begin{array}{cc}
0 & {\rm q}_{+a} \\ {\rm p}_{+a} & 0
\end{array}
\right)\,,\quad
q_{-}^a=\left(
\begin{array}{cc}
0 & {\rm q}_{-}^a \\ {\rm p}_{-}^a & 0
\end{array}
\right).
\ee
The corresponding part of the classical action (\ref{S2-bifund-2}) is given by
\bea
S_{\rm off-diag}&=&-\frac12\int d^3x d^4\theta\big[
\bar{\rm v}_L(\hat\square_{\rm v}+M^2){\rm v}_L
+\bar{\rm v}_R(\hat\square_{\rm v}+M^2){\rm v}_R\nn\\&&
+\bar{\rm q}_+^a{\rm q}_{+a}
+\bar{\rm q}_{-a}{\rm q}_-^a
+\bar{\rm p}_+^a{\rm p}_{+a}
+\bar{\rm p}_{-a}{\rm p}_-^a\nn\\&&
+g{\bf Q}_{+a}(\bar{\rm p}_+^a \bar{\rm v}_L-\bar{\rm q}_+^a{\rm v}_R)
+g\bar{\bf Q}_+^a({\rm p}_{+a}{\rm v}_L-{\rm q}_{+a}\bar{\rm v}_R)\nn\\&&
+g{\bf Q}_{-}^a(\bar{\rm p}_{-a}\bar{\rm v}_R-\bar{\rm q}_{-a}{\rm v}_L)
+g\bar{\bf Q}_{-a}({\rm p}_-^a{\rm v}_R-{\rm q}_-^a\bar{\rm v}_L)
\big]\nn\\&&
-\lambda\left[\int d^3x d^2\theta\,
\varepsilon^{ab}\varepsilon_{cd}{\bf Q}_{+a}{\bf Q}_{-}^c
({\rm q}_{+b}{\rm p}_{-}^d-{\rm p}_{+b}{\rm q}_{-}^d)+c.c.\right].
\label{S-off2}
\eea
It is straightforward to compute the matrix of second
variational derivatives of this action with respect to the quantum
superfields and to find the trance of its logarithm.
The details of this procedure are collected in the Appendix
\ref{Ap3}. Here we present only the result,
\be
\Gamma_{\rm off-diag}=2i{\rm Tr}_+\ln\hat\square_+
+2i{\rm Tr}_+\ln(\hat\square_++{\bf Q}^4)
+2i{\rm Tr_v}\ln \hat \square_{\rm v}
+2i{\rm Tr_v}\ln\left(
1-\frac{M^2}{8\hat\square_{\rm v}}{\cal D}^\alpha\bar{\cal D}^2{\cal
D}_\alpha
\right),
\label{4.41}
\ee
where $M^2$ is given in (\ref{4.37}) and ${\bf Q}^4$ is
\be
{\bf Q}^4=4\lambda\bar\lambda(\bar{\bf Q}_+^a {\bf Q}_{+a})({\bf Q}_-^b\bar {\bf
Q}_{-b})\,.
\ee
The traces of the operators in (\ref{4.41}) can be
computed in a standard way,
\bea
\Gamma_{\rm off-diag}
&=&\frac 1\pi\int d^3x d^4\theta\left[
3{\bf G}\ln{\bf G}+2\sqrt{{\bf G}^2+g^2\bar{\bf Q}_+^a{\bf Q}_{+a}}-2{\bf G}\ln({\bf G}
+\sqrt{{\bf G}^2+g^2\bar{\bf Q}_+^a{\bf Q}_{+a}})\right]\nn\\&&
+\frac1\pi\int d^3x d^4\theta\left[
{\bf G}\ln({\bf G}+\sqrt{{\bf G}^2+{\bf Q}^4})
-\sqrt{{\bf G}^2+{\bf Q}^4}
\right].
\label{G-off-2}
\eea

The contribution from the ghost superfields is given by
(\ref{G-gh}) because the ghost superfield sector is the
same as in the model with one hypermultiplet. With the
contribution from the ghosts, the expressions (\ref{G-diag-2}) and
(\ref{G-off-2}) give together the low-energy effective action for
the model (\ref{S2-bifund-2}),
\bea
\Gamma
&=&\frac 1\pi\int d^3x d^4\theta[
\frac{\sqrt2g}2\sqrt{\bar{\bf Q}_+^a{\bf Q}_{+a}}
+2\sqrt{{\bf G}^2+g^2\bar{\bf Q}_+^a{\bf Q}_{+a}}
-2{\bf G}\ln({\bf G}
+\sqrt{{\bf G}^2+g^2\bar{\bf Q}_+^a{\bf Q}_{+a}})]\nn\\&&
+\frac1\pi\int d^3x d^4\theta\left[
{\bf G}\ln({\bf G}+\sqrt{{\bf G}^2+{\bf Q}^4})
-\sqrt{{\bf G}^2+{\bf Q}^4}
\right].
\label{G-final}
\eea

Consider the asymptotics of the action (\ref{G-final}) at large
values of the gauge coupling constant $g$, \bea \Gamma&=&
\frac1\pi\int d^3x d^4\theta\left[ {\bf G}\ln({\bf G}+\sqrt{{\bf
G}^2+{\bf Q}^4}) -\sqrt{{\bf G}^2+{\bf Q}^4} -{\bf G}\ln(\bar{\bf
Q}_+^a{\bf Q}_{+a}) \right]\nn\\&& +g\frac{4+\sqrt2}{2\pi}\int d^3x
d^4\theta \sqrt{\bar{\bf Q}_+^a{\bf Q}_{+a}} +O(1/g)\,.
\label{G-final1} \eea We point out that the terms in the first line
here are scale invariant sine they have no dependence on $g$.
Moreover, it is easy to see that, up to obvious rescaling of fields,
this part of the effective action coincides with the dual action of
the ABJM theory (\ref{ABJM-dual}). Therefore we conclude that the
model (\ref{S2-bifund-2}) reproduces the ABJM theory as a scale
independent part of its low-energy effective action. This confirms
the statement that the D2 brane probing the tip of the conifold
defined by the constraint (\ref{iii}) and (\ref{fff}) flows to its
infrared superconformal point in which it is dual to the M2 brane
\cite{Aganagic}. However, the terms in the second line in
(\ref{G-final1}) which have the form of the K\"ahler potential are
non-conformal because of the dimensionfull coupling constant $g$.
Moreover, they dominate in the infrared, at large coupling constant,
and the M2 brane dynamics should be suppressed and invisible on this
background. We expect that the higher-loop corrections to the
K\"ahler potential in this model cannot completely cancel the terms
in the second line of (\ref{G-final1}) and do not change the present
conclusions significantly.

One of the natural ways to prevent the appearance of the K\"ahler
potential is to modify the model (\ref{N2-bifund2}) by increasing
the number of supersymmetries at least up to $\cN=3$. In
particular, one can study the effective action in the
Kachru-Silverstein model \cite{Kachru} reduced to three dimensions.
In the three-dimensional case it has $\cN=4$ supersymmetry and quantum
corrections of the form of the K\"ahler potential are forbidden
by the non-renormalization theorems \cite{deBoer,Mirror,AHKSS}.
Alternatively, one can study the effective action in the ABJM
model deformed by the SYM kinetic terms such that it has
$\cN=3$ supersymmetry \cite{BKT}. This problem deserves separate
considerations.

\setcounter{equation}{0}
\section{Summary and discussion}
It is well-known that the 2-branes in M-theory appear in the
strong coupling limit for the D2 branes in the IIA supergravity.
One of the aims of the present paper is to observe the
consequences of this relation between the M2 and D2 branes for the
corresponding field theories living in the world-volume of these branes.
This appears to be possible thanks to the recent progress in
constructing the supergauge theory describing multiple M2 branes
which are known as the BLG and ABJM models \cite{BLG,ABJM}. On the
side of the D2 brane, one has three-dimensional SYM theory which can be
explored by standard methods of quantum field theory.

In the present paper we study the low-energy effective action in the
$\cN=2$ quiver SYM theory with four chiral superfields in the
bifundamental representation and with scale invariant
superpotential. It is known that this model describes the D2 brane
probing the singularity of the conifold $X_3$ fibred over real line
\cite{Aganagic,KT}. As is explained in \cite{Aganagic}, the D2 brane
on such a background considered at strong coupling should reproduce
the M2 brane on the background $X_4={\mathbb C}^4/{\mathbb Z}_k$
which is described by the ABJM theory \cite{ABJM}. In the present
paper we demonstrate that the scale invariant part of the low-energy
effective action of this $\cN=2$ SYM-matter theory precisely
reproduces the classical action of the Abelian ABJM theory,
rewritten in its dual form when one of the scalar superfields is
dualized into a dynamical gauge superfield. However, we observe that,
apart from these scale invariant contributions, the one-loop
effective action receives the non-scale-invariant ones as well. These
non-scale-invariant terms have the form of the K\"ahler potential
and depend linearly on the gauge coupling constant $g$. Therefore
they dominate in the infrared, at large values of the coupling, and
the scale-invariant contributions having the form of the dualized
ABJM action are suppressed by these non-conformal ones. Because of
this phenomenon the correspondence between the D2 and M2 branes
cannot be precisely confirmed.

We guess that the correspondence between the low-energy effective
action of the three-dimensional SYM-matter theory and the classical
action of the ABJM model can be checked more precisely if one
modifies the $\cN=2$ quiver gauge theory (\ref{N2-bifund2}) in such
a way that the K\"ahler potential for chiral superfields would not
appear in the effective action. For instance, one can consider an
analog of the model (\ref{N2-bifund2}) with the $\cN=3$ supersymmetry
\cite{BKT} which prohibits the generation of the K\"ahler potential.
The $\cN=3$, $d=3$ harmonic superspace methods \cite{my} should
be helpful for these considerations. Alternatively, one can study
the effective action in the Kachru-Silverstein model \cite{Kachru}
reduced to three dimensions such that it has the $\cN=4$
supersymmetry. Quantum aspects in these models will be explored
elsewhere.

With the same motivations it would be
interesting to do direct quantum computations of the
low-energy effective actions in many other $\cN=2$
Chern-Simons-matter models which were considered in \cite{JY} from
the point of view of mirror symmetry.

In the present paper we computed also the one-loop effective
actions in the pure SYM models with $\cN=2$, $\cN=4$ and $\cN=8$
supersymmetry. The low-energy effective action in the $\cN=2$ and
$\cN=4$ SYM starts with the terms which contain $F^2$
in components while the $\cN=8$ SYM effective action starts from
$F^4$. The generation of the $F^2$ term by quantum corrections in the
$\cN=8$ SYM is forbidden by the non-renormalization theorem
\cite{DS} which is analogous to the one in the $\cN=4$, $d=4$ SYM. The
leading $F^2$ terms in the $\cN=2$ and $\cN=4$ SYM were derived
originally by employing the mirror symmetry
\cite{deBoer,deBoer1,Mirror,AHKSS}. We stress that by doing
direct quantum computations in the $\cN=2$, $d=3$ superspace we
derive non only these leading $F^2$ terms, but also all higher order
ones giving $F^{2n}$ in components for all positive integer $n$.

An interesting feature of the $\cN=4$ SYM model is that its
low-energy effective action starts from the terms which are
superconformal and coincide with the classical action of the
Abelian Gaiotto-Witten model \cite{GW} written in its dualized
form \cite{KLL}. A similar feature was mentioned in \cite{BPS} for the
hypermultiplet interacting with the Abelian background $\cN=4$
gauge multiplet. This can be considered as a prototype of the
correspondence between the effective actions of gauge theories on
D2 and M2 branes, but living in the reduced space-time.

A natural continuation of the present work is the study the
low-energy effective action in the ABJM model deformed by the SYM
kinetic terms for the gauge superfields such that the
supersymmetry is reduced down to $\cN=3$ \cite{BKT}. When the
gauge coupling is sent to infinity and the SYM kinetic terms drop
out, the $\cN=6$ supersymmetry is restored and this should give
us the low-energy effective action in the ABJM theory. This will
result in the superspace analogs of the results of similar
component computations which were done in \cite{quantABJM}. It is
interesting also to investigate the effective action in other superconformal
three-dimensional gauge theories, considered e.g.\ in \cite{Deform},
which are interesting from the point of view of the AdS/CFT
correspondence.

Another tempting problem is the study of the two-loop quantum
contributions to the effective action in the $\cN=2$ supersymmetric
electrodynamics both in the Coulomb and in the Higgs branches. The
one-loop effective action on the Coulomb branch in this model was
considered in \cite{BPS} while the one-loop corrections to the Higgs
branch were partly considered in the present work, e.g., from
(\ref{G-diag}). Note that these one-loop results are not new because
they were obtained quite a while by utilizing the power of the
mirror symmetry \cite{deBoer,Mirror,AHKSS}. However, two-loop
corrections to these effective actions are promising. In particular, the
two-loop K\"ahler potential in the $\cN=1$ supersymmetric
electrodynamics was studied in \cite{daSilva}, but similar $\cN=2$
superspace considerations are welcome.

\vspace{3mm}
{\bf Acknowledgments}\\[3mm]
We are grateful to S.J. Gates and S.M. Kuzenko for drawing our
attention to some important references.
I.B.S.~is indebted to D. Belyaev, O. Lechtenfeld and D. Sorokin
for useful discussions. The work was partially supported by RFBR grant,
project No 09-02-00078 and by the grant for LRSS, project No
3558.2010.2. I.L.B. and I.B.S. acknowledge the
support from the RFBR grants No 10-02-90446 and No 09-02-91349 as
well as from a DFG grant, project No 436 RUS/113/669. N.G.P.\
acknowledges the support from RFBR grant, project No 08-02-00334.
The work of I.B.S. was supported by the Marie Curie research
fellowship No 236231, ``QuantumSupersymmetry'' during his work at
INFN, Padova and by the fellowship of the Dynasty foundation as
well as by the RF Federal programm ``Kadry'' contract Nr P691
during the work at Tomsk Polytechnic University.

\appendix
\section{Dualization of the Gaiotto-Witten and ABJM models}
\setcounter{equation}{0}
\renewcommand{\theequation}{A.\arabic{equation}}
\subsection{Dualization of the Abelian Gaiotto-Witten model}
Consider the following model in the $\cN=2$ superspace,
\be
S_{\rm GW}=-\int d^3x d^4\theta[
\bar Q_+ e^{2V}Q_++\bar Q_- e^{-2V}Q_-+4\hat V G]\,,
\label{SGW}
\ee
where $(\bar Q_+,Q_-)$ is the hypermultiplet while $V$ and $\hat V$ are
two gauge superfields with the corresponding field strengths $G$
and $\hat G$.
The action (\ref{SGW}) is invariant under the following hidden
$\cN=2$ supersymmetry
\bea
\delta Q_{+}&=&\frac12\bar D^2(\bar\theta^\alpha\bar\epsilon_\alpha
 \bar Q_- e^{- 2V})\,,\qquad
\delta Q_{-}=-\frac12\bar D^2(\bar\theta^\alpha\bar\epsilon_\alpha
 \bar Q_+ e^{ 2V})\,,\nn\\
\delta\hat V&=&-i\epsilon^\alpha\bar\theta_\alpha Q_+Q_-
-i\bar\epsilon^\alpha\theta_\alpha \bar Q_+\bar Q_-\,,\qquad
\delta V=0\,.
\eea
Here $\epsilon_\alpha$ is the complex parameter of the hidden
supersymmetry.

It should be noted that a four-dimensional analog of
the action (\ref{SGW}) was considered originally in \cite{LR} as a dual
form of the improved $\cN=2$, $d=4$ tensor multiplet action.
Alternatively, the action (\ref{SGW}) can be viewed as a BF-type theory
coupled to matter chiral superfields. The importance of
three-dimensional
supersymmetric BF-type theories was pointed out in \cite{Gates}, where the
authors studied the coupling of such models to supergravity as
well as some aspects of their duality and mirror symmetry. More
recently, in \cite{KLL} it was shown that (\ref{SGW}) corresponds to the
classical action of the Abelian Gaiotto-Witten model \cite{GW} which
is, in general, a Chern-Simons-matter theory with N = 4
supersymmetry.

Following \cite{LR}, we consider the equation of motion
for the gauge superfield $V$,
\be
\bar Q_+ e^{2V}Q_+-\bar Q_-e^{-2V}Q_- +2\hat G=0\,.
\ee
It is solved by
\be
e^{-2V}=\frac{\hat G+\sqrt{\hat G^2+\bar Q_+Q_+\bar Q_- Q_-}}{\bar Q_-
Q_-}\,,
\ee
or
\be
V=-\frac12\ln(\hat G+\sqrt{\hat G^2+\bar Q_+Q_+\bar Q_- Q_-})
+\frac12 \ln(\bar Q_- Q_-)\,.
\ee
It is convenient to denote
\be
\bar Q_+\bar Q_-=\bar \Phi\,,\qquad
Q_+ Q_-=\Phi\,.
\ee
Substituting this solution back into the action (\ref{SGW}) we get
\be
\tilde S_{\rm GW}=2\int d^3x d^4\theta\left[
\hat G\ln(\hat G+\sqrt{\hat G^2+\bar\Phi\Phi})
-\sqrt{\hat G^2+\bar \Phi\Phi}\right]\,.
\label{GW-dual}
\ee
This action is known to be $\cN=4$ supersymmetric and
superconformal, see, e.g., \cite{HKLR}.

In (\ref{GW-dual}) both the gauge superfield $\hat V$
and the chiral superfield $\Phi$ are propagating while in the
original action (\ref{SGW}) the gauge superfields are non-dynamical
and only chiral superfields $Q_\pm$ propagate. Hence, one of the
chiral superfields in (\ref{SGW}) is dualized into the Abelian
gauge superfield with the preservation of the supersymmetry and
superconformal invariance. Therefore we refer to (\ref{GW-dual})
as a dual representation of the Abelian action for
Gaiotto-Witten theory, although (\ref{GW-dual}) was known long before
within the study of dualities among the tensor multiplet and
supersymmetric sigma models \cite{HKLR,LR}.

\subsection{Dualization of the Abelian ABJM model}
The ABJM model \cite{ABJM} is similar to the Gaiotto-Witten theory, but the
hypermultiplet is a SU(2) doublet, $(\bar Q_+^a,Q_-^a)$, $a=1,2$. The action
in the Abelian case is quite simple,
\be
S_{\rm ABJM}=-\int d^3x d^4\theta[
\bar Q_+^a e^{2V}Q_{+a}
+ Q_-^a e^{-2V}\bar Q_{-a}+4\hat VG
]\,.
\label{S-ABJM}
\ee
Analogous four-dimensional models were considered in \cite{LR}
within the study of non-linear sigma-models.

The action (\ref{S-ABJM}) possesses hidden $\cN=4$ supersymmetry,
\bea
\delta Q_{+a}&=&\frac12\bar D^2[(\bar\theta^\alpha\epsilon_{\alpha\, a}{}^b)
\bar Q_{-b}e^{-2V}]\,,\quad
\delta Q_{-}^a=-\frac12\bar D^2[(\bar\theta^\alpha \epsilon_\alpha{}^a{}_b)
\bar Q_{+}^b e^{2V}]\,,\nn\\
\delta\hat V&=&-i\bar\theta^\alpha\epsilon_{\alpha}{}^a{}_b \ Q_{+a}
Q_-^b
-i\theta^\alpha\epsilon_{\alpha\,a}{}^b \bar Q_+^a \bar
Q_{-b}\,,\quad
\delta V=0\,.
\label{N6}
\eea
Here $\epsilon^{\alpha\,a}{}_{b}$ is a real supersymmetry parameter
which bears two SU(2) indices. Its indices are raised
or lowered with the $\varepsilon^{ab}$ symbol. The reality of this
supersymmetry parameter is required by the on-shell closure of the
supersymmetry transformations. Together with the explicit $\cN=2$
supersymmetry, the transformations (\ref{N6}) form the $\cN=6$
supersymmetry of the ABJM theory.

The equation of motion for the gauge superfield $V$
\be
\bar Q_+^a e^{2V}Q_{+a}-\bar Q_{-a}e^{-2V}Q_-^a +2\hat G=0
\ee
is solved by
\be
e^{-2V}=\frac{\hat G+\sqrt{\hat G^2+\bar Q_+^aQ_{+a}\bar Q_{-b} Q_-^b}}{
\bar Q_{-a} Q_-^a}\,,
\ee
or
\be
V=-\frac12\ln(\hat G+\sqrt{\hat G^2+\bar Q_+^aQ_{+a}\bar Q_{-b} Q_-^b})
+\frac12 \ln(\bar Q_{-a} Q_-^a)\,.
\ee
Substituting this solution back into (\ref{S-ABJM}) we
get the dual representation of the ABJM action,
\bea
\tilde S_{\rm ABJM}
&=&2\int d^3x d^4\theta\bigg[
\hat G\ln(\hat G+\sqrt{\hat G^2+\bar Q_+^a Q_{+a}Q_{-b}\bar Q_-^b})
-\sqrt{\hat G^2+\bar Q_+^a Q_{+a}Q_{-b}\bar Q_-^b}\nn\\
&&-\hat G\ln(Q_-^a\bar Q_{-a})
\bigg].
\label{ABJM-dual}
\eea
In components, this action contains supersymmetric and
superconformal generalization of the Maxwell $F^2$ term which
originates from the superfield strength $\hat G$

At first sight the dual ABJM action (\ref{ABJM-dual}) contains four propagating
chiral superfields $Q_{+a}$ and $Q_-^a$ as well as the propagating
gauge superfield $\hat G$. However, only three of four chiral
superfields are independent here. Indeed, one can redefine the
chiral superfields such that the SU(2) invariance becomes implicit,
\be
\Phi_1=Q_{+1} Q_-^1\,,\quad
\Phi_2=Q_{+2} Q_-^2\,,\quad
\Phi_3=Q_{+2} Q_-^1\,.
\ee
In terms of these superfields the action (\ref{ABJM-dual}) reads
\bea
\tilde S_{\rm ABJM}&=&2\int d^3x d^4\theta\bigg[
\hat G\ln\left(\hat G+\sqrt{\hat G^2+\bar\Phi^i\Phi_i+\frac{\Phi_1\bar\Phi^1\Phi_2\bar\Phi^2
}{\Phi_3\bar\Phi^3}}\right)\nn\\&&
-\sqrt{\hat G^2+\bar\Phi^i\Phi_i+\frac{\Phi_1\bar\Phi^1\Phi_2\bar\Phi^2
}{\Phi_3\bar\Phi^3}}
-\hat G\ln(1+\frac{\bar\Phi^2\Phi_2}{\bar\Phi^3\Phi_3})
\bigg]\,.
\label{ABJM-dual-1}
\eea
In this action one can see only three propagating chiral superfields $\Phi_i$, $i=1,2,3$,
which, together with the gauge superfield $\hat G$
give correct number of degrees of freedom of the ABJM theory with
the action (\ref{S-ABJM}).

\section{Some technical details of one-loop computations}
\setcounter{equation}{0}
\renewcommand{\theequation}{B.\arabic{equation}}
\subsection{Transformation of the effective action in the model (\ref{S-diag})
to the form (\ref{4.25})}
\label{ApB1}
The one-loop effective action is given by the standard expression
$\Gamma=\frac i2\Tr\ln H$, where $H$ is the matrix of second
variational derivatives of the action (\ref{S-diag}) with respect
to the quantum superfields. It has the following block form
\be
H=\left(
\begin{array}{cc}
A & B \\ C & D
\end{array}
\right)=
\left(
\begin{array}{cc}
\frac{\delta^2 S}{\delta{\rm v}_{L,R}(z)\delta{\rm
v}_{L,R}(z')} & \frac{\delta^2 S}{\delta{\rm v}_{L,R}(z)\delta{\rm
q}_\pm(z')} \\ \frac{\delta^2 S}{\delta{\rm
q}_\pm(z)\delta{\rm v}_{L,R}(z')} & \frac{\delta^2 S}{\delta{\rm q}_\pm(z)\delta{\rm
q}_\pm(z')}
\end{array}
\right),
\label{H}
\ee
where the block matrices are
\bea
A&=&-\left(
\begin{array}{cc}
\square+M^2 & -M^2\\ -M^2 & \square+M^2
\end{array}
\right)\delta^7(z-z')\,,\nn\\
B&=&-\frac g2\left(
\begin{array}{cccc}
\bar {\bf Q}_+\delta_+(z,z') & -\bar{\bf Q}_-\delta_+(z,z')&
{\bf Q}_+\delta_-(z,z') & -{\bf Q}_-\delta_-(z,z')\\
-\bar {\bf Q}_+\delta_+(z,z') & \bar{\bf Q}_-\delta_+(z,z')&
-{\bf Q}_+\delta_-(z,z') & {\bf Q}_-\delta_-(z,z')
\end{array}
\right),
\nn\\
C&=&B^\dag\nn\\
D&=&\left(
\begin{array}{cccc}
0 & 0 & \frac18\bar D^2\delta_-(z,z') & 0\\
0 & 0 & 0 & \frac18\bar D^2\delta_-(z,z')\\
\frac18D^2\delta_+(z,z') &0 &0 &0\\
0 & \frac18 D^2\delta_+(z,z') &0&0
\end{array}
\right).
\label{D}
\eea
Using standard decomposition for the block matrices,
\be
\left(
\begin{array}{cc}
A&B\\C&D
\end{array}
\right)=
\left(
\begin{array}{cc}
1&BD^{-1}\\0&1
\end{array}
\right)
\left(
\begin{array}{cc}
A-BD^{-1}C&0\\0&D
\end{array}
\right)
\left(
\begin{array}{cc}
1&0\\D^{-1}C&1
\end{array}
\right),
\label{mat-rel}
\ee
we get the following representation for the one-loop effective
action
\bea
\Gamma_{\rm diag}&=&\frac i2 {\rm Tr_v}\ln
\left(
\begin{array}{cc}
-\square-M^2+\frac1{16}\frac{M^2}\square\{D^2,\bar D^2 \} &
M^2-\frac1{16}\frac{M^2}\square\{D^2,\bar D^2 \} \\
M^2-\frac1{16}\frac{M^2}\square\{D^2,\bar D^2 \} &
-\square-M^2+\frac1{16}\frac{M^2}\square\{D^2,\bar D^2 \}
\end{array}
\right)\nn\\
&=&\frac i2 {\rm Tr_v}\ln
\left(
\begin{array}{cc}
-\square+\frac18\frac{M^2}\square D^\alpha\bar D^2 D_\alpha &
-\frac18\frac{M^2}\square D^\alpha \bar D^2 D_\alpha \\
-\frac18\frac{M^2}\square D^\alpha\bar D^2 D_\alpha&
-\square+\frac18\frac{M^2}\square D^\alpha\bar D^2 D_\alpha
\end{array}
\right)\nn\\
&=&\frac i2 {\rm Tr_v}\ln(1-\frac 14\frac{M^2}{\square^2}D^\alpha\bar D^2 D_\alpha)
=-\frac i{16}{\rm Tr_v}\frac1\square\ln(1+\frac{2M^2}{\square})D^\alpha
\bar D^2 D_\alpha\,.
\label{b8}
\eea
Here we have applied the standard identity for the covariant spinor
derivatives,
\be
\frac1{16}\{D^2,\bar D^2\} -\frac18D^\alpha \bar D^2 D_\alpha
=\square\,.
\ee
All operators in (\ref{b8}) act in the full superspace.

A shorter way to get the representation (\ref{b8}) for the
effective action is to use the Landau gauge instead of the
Fermi-Feynman one. In this case there are no mixed
contributions with quantum vector and matter propagators and the
last two lines in (\ref{S2-bifund}) can be simply omitted.

\subsection{Transformation of the effective action in the model (\ref{S-off})
to the form (\ref{4.30})}

\label{ApB2}
The matrix of second variational derivatives of the
action (\ref{S-off}) with respect to the quantum superfields
has the structure (\ref{H}), but with the block matrices given by
\bea
\label{A}
A&=&-\frac12\left(
\begin{array}{cccc}
0& 0& \hat\square_{\rm v}+M^2 &0\\
0& 0& 0 &\hat\square_{\rm v}+M^2\\
\hat\square_{\rm v}+M^2 &0&0&0\\
0&\hat\square_{\rm v}+M^2&0&0
\end{array}
\right)\delta^7(z-z'),\\
B&=&-\frac g2\left({\small
\begin{array}{cccccccc}
0& \bar{\bf Q}_+\hat\delta_{+}& 0&0&0&0& -{\bf Q}_-\hat\delta_{-}&0\\
0&0&0& \bar{\bf Q}_-\hat\delta_{+} & -{\bf Q}_+\hat\delta_{-} &0&0&0\\
0&0&-\bar{\bf Q}_-\hat\delta_{+} &0&0& {\bf Q}_+\hat\delta_{-} &0&0\\
-\bar{\bf Q}_+\hat\delta_{+}&0&0&0&0&0&0&{\bf Q}_-\hat\delta_{-}
\end{array}}
\right),\nn\\
C&=&B^\dag\,,\nn
\eea
\bea
D&=&\left({\small
\begin{array}{cccccccc}
0&0&0&0& \frac18\bar{\cal D}^2\hat\delta_{-} &0&0&0\\
0&0&0&0& 0& \frac18\bar{\cal D}^2\hat\delta_{-}&0&0\\
0&0&0&0& 0& 0& \frac18\bar{\cal D}^2\hat\delta_{-}&0\\
0&0&0&0& 0& 0& 0& \frac18\bar{\cal D}^2\hat\delta_{-}\\
\frac18{\cal D}^2\hat\delta_{+} &0&0&0 & 0&0&0&0\\
0& \frac18{\cal D}^2\hat\delta_{+} &0&0&0&0&0&0\\
0& 0&\frac18{\cal D}^2\hat\delta_{+}&0&0&0&0&0\\
0&0&0& \frac18{\cal D}^2\hat\delta_{+}&0&0&0&0
\end{array}}
\right).\nn
\eea
Here $\hat\delta_{\pm}\equiv\hat\delta_{\pm}(z,z')$
are the covariantly
(anti)chiral delta functions (in contrast to the flat ones
$\delta_\pm(z,z')$ which were used (\ref{D})),
\be
\hat\delta_{+}(z,z')=-\frac14\bar{\cal D}^2\delta^7(z-z')\,,\qquad
\hat\delta_{-}(z,z')=-\frac14{\cal D}^2\delta^7(z-z')\,.
\ee

The background superfields ${\bf Q}_\pm$ are constant,
$D_\alpha {\bf Q}_\pm=0$, since we are interested in the low-energy
effective action in the constant field
approximation. The background covariant spinor derivative
${\cal D}_\alpha$ is defined as in the Abelian case,
${\cal D}_\alpha=D_\alpha+{\bf V}_\alpha$, where the gauge
connection ${\bf V}_\alpha$ acts just by multiplication. Therefore
the covariant spinor derivatives do not hit the background (anti)chiral
superfields,
${\cal D}_\alpha {\bf Q}_\pm X={\bf Q}_\pm {\cal D}_\alpha X$.
Keeping this in mind, we get the following form for the matrix
$A-BD^{-1}C$,
\be
A-BD^{-1}C=\left(
\begin{array}{cc}
0& E\\ E&0
\end{array}
\right),
\label{A-BDC}
\ee
where
\be
E=-\frac12\left(
\begin{array}{cc}
\hat\square_{\rm v}+M^2-\frac{M^2}{16}\{{\cal D}^2,\bar{\cal D}^2\}
\frac1{\hat\square_{\rm v}} &0\\0&
\hat\square_{\rm v}+M^2-\frac{M^2}{16}\{{\cal D}^2,\bar{\cal D}^2\}
\frac1{\hat\square_{\rm v}}
\end{array}
\right)\delta^7(z-z')\,.
\label{E}
\ee
Here we have used standard relations between the covariant
derivatives and covariant box operators \cite{BPS},
\be
{\cal D}^2\hat \square_+={\cal D}^2\hat \square_{\rm v}
=\hat\square_{\rm v}{\cal D}^2\,,\qquad
\bar{\cal D}^2\hat\square_-=\bar{\cal D}^2\hat\square_{\rm v}=
\hat\square_{\rm v}\bar{\cal D}^2\,.
\label{id}
\ee

Consider now the expression
$\hat\square_{\rm v}+M^2-\frac{M^2}{16}\{{\cal D}^2,\bar{\cal D}^2\}
\frac1{\hat\square_{\rm v}}\delta^7(z-z')$ which appears in
(\ref{E}). Using the identity (\ref{v-square}), it can be
rewritten as
\be
\hat\square_{\rm v}+\frac{M^2}{\hat\square_{\rm v}}
\left[\frac1{16}\{{\cal D}^2,\bar{\cal D}^2\}-\frac18{\cal D}^\alpha\bar{\cal D}^2
{\cal D}_\alpha+\frac i2({\cal D^\alpha{\bf W}_\alpha})+i{\bf W}^\alpha {\cal
D}_\alpha\right]
-\frac{M^2}{16}\{{\cal D}^2,\bar{\cal D}^2\}
\frac1{\hat\square_{\rm v}}\,.
\label{B12}
\ee
As soon as this expression is considered under the integral over
the full superspace, we can omit the superfield strength ${\bf W}_\alpha$
since it is responsible for the higher-order contributions to the
effective action (see  (\ref{Q-constr1}) and comments nearby).
As a result, (\ref{B12}) simplifies,
\be
\hat\square_{\rm v}-\frac18\frac{M^2}{\hat\square_{\rm v}}
{\cal D}^\alpha\bar{\cal D}^2
{\cal D}_\alpha\,.
\label{B13}
\ee
We point out that the operator $\hat\square_{\rm
v}$ commutes with the covariant spinor
derivatives ${\cal D}_\alpha$ and $\bar{\cal D}_\alpha$ because,
according to (\ref{alg2}), these commutators are proportional to
the spinor superfield strengths which we neglect in the approximation
(\ref{Q-constr1}).

Applying the standard matrix relation (\ref{mat-rel}), we rewrite
the one-loop effective action in the model (\ref{S-off}) in the
following form
\bea
\Gamma_{\rm off-diag}&=&\frac i2 \Tr\ln H=\frac i2\Tr\ln (A-BD^{-1}C)+\frac
i2\Tr\ln D\nn\\&=&
2i{\rm Tr_v}\ln\left( \hat\square_{\rm v}-\frac18\frac{M^2}{\hat\square_{\rm v}}
{\cal D}^\alpha\bar{\cal D}^2
{\cal D}_\alpha\right)
+2i{\rm Tr}_+\ln \hat\square_+\nn\\
&=&2i{\rm Tr}_+\hat\square_++2i{\rm Tr_v}\ln\hat\square_{\rm v}
+2i{\rm Tr_v}\ln\left(
1-\frac{M^2}{8\hat\square_{\rm v}^2}{\cal D}^\alpha\bar{\cal
D}^2{\cal D}_\alpha
\right).
\label{B14}
\eea
Our final comments concern the computations of the last term in
(\ref{B14}). For the considered background, when we neglect the
superfield strengths ${\bf W}_\alpha$ and $\bar{\bf W}_\alpha$ in
the full superspace, it is sufficient to consider only the
following terms in $\hat\square_{\rm v}$,
\be
\hat\square_{\rm v}\approx{\cal D}^m{\cal D}_m+{\bf G}^2\,.
\ee
Then we can use the relation $({\cal D}^\alpha\bar{\cal D}^2{\cal D}_\alpha)^n
=(-8\hat\square_{\rm v})^{n-1}{\cal D}^\alpha\bar{\cal D}^2{\cal
D}_\alpha$ to get
\bea
&&2i{\rm Tr_v}\ln\left(
1-\frac{M^2}{8\hat\square_{\rm v}^2}{\cal D}^\alpha\bar{\cal
D}^2{\cal D}_\alpha
\right)\delta^7(z-z')=-\frac i4{\rm Tr_v}
\frac1{\hat\square_{\rm v}}\ln(1+\frac{M^2}{\hat\square_{\rm v}})
{\cal D}^\alpha\bar{\cal D}^2{\cal
D}_\alpha \delta^7(z-z')\nn\\
&&=-4i\Tr\frac1{\square+{\bf G}^2}\ln(1+\frac1{\square+{\bf G}^2})
\delta^3(x-x')\,.
\label{B15}
\eea
Here we have applied also the first identity from (\ref{4.26}).

\subsection{Transformation of the effective action in the model (\ref{S-off2})
to the form (\ref{4.41})}
\label{Ap3}
As usual, the one-loop effective action is defined by the matrix
of second variational derivatives (\ref{H})
with respect to the quantum superfields. In the case of the
action (\ref{S-off2}) the block matrix $A$  in (\ref{H})
is the same as (\ref{A}), but the matrices $B$, $C$ and $D$ are
now
\bea
B&=&-\frac g2\left({\scriptsize
\begin{array}{cccccccc}
0& \bar{\bf Q}_+^a\hat\delta_{+}& 0&0&0&0& -{\bf Q}_-^a\hat\delta_{-}&0\\
0&0&0& \bar{\bf Q}_{-a}\hat\delta_{+} & -{\bf Q}_{+a}\hat\delta_{-} &0&0&0\\
0&0&-\bar{\bf Q}_{-a}\hat\delta_{+} &0&0& {\bf Q}_{+a}\hat\delta_{-} &0&0\\
-\bar{\bf Q}_+^a\hat\delta_{+}&0&0&0&0&0&0&{\bf Q}_-^a\hat\delta_{-}
\end{array}}
\right),\\
C&=&B^\dag\,,\nn\\
D&=&\left({\scriptsize
\begin{array}{cccccccc}
0&0&0&-N^b{}_a\hat\delta_+& \frac18\delta_a^b\bar{\cal D}^2\hat\delta_{-} &0&0&0\\
0&0&N^b{}_a\hat\delta_+&0& 0& \frac18\delta_a^b\bar{\cal D}^2\hat\delta_{-}&0&0\\
0&N^b{}_a\hat\delta_+&0&0& 0& 0& \frac18\delta_a^b\bar{\cal D}^2\hat\delta_{-}&0\\
-N^b{}_a\hat\delta_+&0&0&0& 0& 0& 0& \frac18\delta_a^b\bar{\cal D}^2\hat\delta_{-}\\
\frac18\delta_a^b{\cal D}^2\hat\delta_{+} &0&0&0 & 0&0&0&-\bar N^b{}_a\hat\delta_-\\
0& \frac18\delta_a^b{\cal D}^2\hat\delta_{+} &0&0&0&0&\bar N^b{}_a\hat\delta_-&0\\
0& 0&\frac18\delta_a^b{\cal D}^2\hat\delta_{+}&0&0&\bar N^b{}_a\hat\delta_-&0&0\\
0&0&0& \frac18\delta_a^b{\cal D}^2\hat\delta_{+}&-\bar N^b{}_a\hat\delta_-&0&0&0
\end{array}}
\right),\nn
\label{DD}
\eea
where
\be
N^a{}_b=\lambda\varepsilon^{ac}\varepsilon_{bd}{\bf
Q}_{+c}{\bf Q}_-^d\,,\qquad
\bar N^b{}_a=\bar\lambda\varepsilon^{bd}\varepsilon_{ac}
\bar{\bf Q}_+^c\bar{\bf Q}_{-d}\,.
\label{N}
\ee

Let us introduce also the following superfields
\bea
S^a{}_c\equiv N^a{}_b\bar N^b{}_c&=&-\lambda\bar\lambda({\bf Q}_-^e\bar{\bf Q}_{-e})
\varepsilon^{ab}\varepsilon_{cd}{\bf Q}_{+b}\bar{\bf Q}_+^d\,,\nn\\
\bar S^a{}_c\equiv \bar N^a{}_b N^b{}_c&=& -\lambda\bar\lambda(\bar{\bf Q}_+^e{\bf Q}_{+e})
\varepsilon^{ab}\varepsilon_{cd}{\bf Q}_-^d \bar{\bf Q}_{-b}\,.
\label{S}
\eea
They have the following properties
\bea
S^a{}_a&=&{\bf Q}^4\,,\qquad
\bar S^a{}_a={\bf Q}^4\,,\nn\\
\bar S^a{}_c\bar S^c{}_d&=&{\bf Q}^4 \bar S^a{}_d\,,\qquad
S^a{}_c S^c{}_d={\bf Q}^4 S^a{}_d\,,\nn\\
S^n&=&({\bf Q}^4)^{n-1}S\,,\qquad \bar S^n=({\bf Q}^4)^{n-1}\bar S\,,
\label{SS}
\eea
where ${\bf Q}^4$ is
\be
{\bf Q}^4=\lambda\bar\lambda(\bar{\bf Q}_+^a {\bf Q}_{+a})({\bf Q}_-^b\bar {\bf
Q}_{-b})\,.
\label{Q4}
\ee

To begin with, let us compute the trace of the logarithm of the
matrix $D$ given by (\ref{DD}). This matrix consists of four
identical blocs which contribute as follows
\bea
\frac i2\Tr\ln D&=&2i\Tr\ln
\left(
\begin{array}{cc}
-N^b{}_a\hat\delta_+ & \frac18\delta^b_a\bar{\cal D}^2\hat\delta_-\\
\frac18\delta^b_a{\cal D}^2\hat\delta_+ & -\bar N^b{}_a\hat\delta_-
\end{array}
\right)\nn\\
&=&2i\Tr\ln\left(
\begin{array}{cc}
\delta^a_b\hat\square_+\hat\delta_+ &
 -\frac12\bar N^a{}_b\bar{\cal D}^2\hat\delta_- \\
-\frac12N^a{}_b{\cal D}^2\delta_+ &
 \delta^a_b\hat\square_-\hat\delta_-
\end{array}
\right)
-2i\Tr\ln\left(
\begin{array}{cc}
0&\frac12\delta^a_b\bar{\cal D}^2\hat\delta_- \\
\frac12\delta^a_b{\cal D}^2\hat\delta_+ &0
\end{array}
\right)\nn\\
&=&2i{\rm Tr}_+\ln\hat\square_+ +2i{\rm Tr}_+\ln(\hat\square_++4{\bf Q}^4)\,.
\label{B20}
\eea

Now we comment on the computation of the matrix $A-BD^{-1}C$. In
fact, this matrix has the same block structure as (\ref{A-BDC}).
Therefore it is sufficient to compute only one element in it. One
of the non-vanishing elements in the matrix $BD^{-1}C$ is given by the
following expression
\bea
&&\frac{g^2}4(\bar{\bf Q}_{-b}, -{\bf Q}_{+b})
\left(
\begin{array}{cc}
-N^b{}_a\hat\delta_+ & \frac18\delta_a^b\bar{\cal
D}^2\hat\delta_-\\
\frac18\delta_a^b{\cal D}^2\hat\delta_+ &
 -\bar N^b{}_a\hat\delta_-
\end{array}
\right)^{-1}
\left(
\begin{array}c
-\bar{\bf Q}_+^a\\{\bf Q}_-^a
\end{array}
\right)\nn\\
&=&\frac{g^2}4(\bar{\bf Q}_{-b}, -{\bf Q}_{+b})
\left(
\begin{array}{cc}
4\bar N^b{}_c(\delta^c_a+\frac{4S^c{}_a}{\hat\square_+-4{\bf Q}^4})
 \frac1{\hat\square_+}\hat\delta_+ &
\frac12(\delta^b_a+\frac{4\bar S^b{}_a}{\hat\square_+-4{\bf
Q}^3})\frac1{\hat\square_+}\bar{\cal D}^2\hat\delta_- \\
\frac12(\delta^b_a+\frac{4 S^b{}_a}{\hat\square_--4{\bf
Q}^3})\frac1{\hat\square_-}{\cal D}^2\hat\delta_+ &
4 N^b{}_c(\delta^c_a+\frac{4\bar S^c{}_a}{\hat\square_--4{\bf Q}^4})
 \frac1{\hat\square_-}\hat\delta_-
\end{array}
\right)
\left(
\begin{array}c
-\bar{\bf Q}_+^a\\{\bf Q}_-^a
\end{array}
\right)\nn\\
&=&\frac{g^2}8(\bar{\bf Q}_+^a{\bf Q}_{+a})\frac1{\hat\square_-}{\cal
D}^2\hat\delta_+
+\frac{g^2}8({\bf Q}_-^a \bar{\bf Q}_{-a})\frac1{\hat\square_+}\bar{\cal
D}^2\hat\delta_-
=-\frac{M^2}{32}\{{\cal D}^2,\bar{\cal D}^2\}\frac1{\hat\square_{\rm
v}}\delta^7(z-z')\,.
\eea
Here we have used the properties (\ref{SS}) of the matrices
(\ref{N}) and (\ref{S}), the identities (\ref{id}) and the
constraint (\ref{Q-constr-2}). The full matrix $-BD^{-1}C$ reads
\be
-BD^{-1}C=\left(
\begin{array}{cccc}
0&0&1&0\\
0&0&0&1\\
1&0&0&0\\
0&1&0&0
\end{array}
\right)\frac{M^2}{32}\{{\cal D}^2,\bar{\cal D}^2\}\frac1{\hat\square_{\rm
v}}\delta^7(z-z')\,.
\ee
Summing up this matrix with (\ref{A}), we get the matrix
$A-BD^{-1}C$ exactly in the form (\ref{A-BDC},\ref{E}). Hence, the
corresponding contributions to the effective action
can be extracted from (\ref{B14}),
\be
\frac i2\Tr\ln(A-BD^{-1}C)=2i{\rm Tr_v}\ln \hat\square_{\rm v}
+2i{\rm Tr_v}\ln\left(
1-\frac{M^2}{8\hat\square_{\rm v}^2}{\cal D}^\alpha \bar{\cal D}^2
 {\cal D}_\alpha
\right).
\ee
This expression, together with (\ref{B20}), defines the structure
of the one-loop effective action in the model (\ref{S-off2}),
\be
\Gamma_{\rm off-diag}=2i{\rm Tr}_+\ln\hat\square_+
+2i{\rm Tr}_+\ln(\hat\square_++{\bf Q}^4)
+2i{\rm Tr_v}\ln \hat \square_{\rm v}
+2i{\rm Tr_v}\ln\left(
1-\frac{M^2}{8\hat\square_{\rm v}}{\cal D}^\alpha\bar{\cal D}^2{\cal
D}_\alpha
\right).
\ee

\end{document}